\documentclass[aps,prb,showpacs,twocolumn,floats,epsfig]{revtex4}
\usepackage{amssymb}
\usepackage{amsbsy}
\usepackage{amsmath}
\usepackage{epsfig}
\usepackage{color}

\begin{document}

\title{Transport across junctions of a Weyl and a multi-Weyl semimetal}

\author{Debabrata Sinha  and   K. Sengupta}

\affiliation{School of Physical Sciences, Indian Association for the
Cultivation of Science, 2A and 2B Raja S.C. Mullick Road,
Jadavpur-700032, India.}

\date{\today}

\begin{abstract}

We study transport across junctions of a Weyl and a multi-Weyl
semimetal (WSM and a MSM) separated by a region of thickness $d$
which has a barrier potential $U_0$. We show that in the thin
barrier limit ($U_0 \to \infty$ and $d \to 0$ with $\chi=U_0
d/(\hbar v_F)$ kept finite, where $v_F$ is velocity of low-energy
electrons and $\hbar$ is Planck's constant), the tunneling
conductance $G$ across such a junction becomes independent of
$\chi$. We demonstrate that such a barrier independence is a
consequence of the change in the topological winding number of the
Weyl nodes across the junction and point out that it has no analogue
in tunneling conductance of either junctions of two-dimensional
topological materials (such as graphene or topological insulators)
or those made out of WSMs or MSMs with same topological winding
numbers. We study this phenomenon both for normal-barrier-normal
(NBN) and normal-barrier-superconductor (NBS) junctions involving
WSMs and MSMs with arbitrary winding numbers and discuss experiments
which can test our theory.

\end{abstract}

\maketitle

\section{Introduction}
\label{intro}

A Weyl semimetal (WSM) hosts a three-dimensional (3D) gapless
topological state whose wavefunction carries a non-zero topological
winding number arising out of singularity in ${\bf k}$ space
\cite{weylrev, ashvin1, ybk1,exp1}. These singularities occur at
Weyl points where the conduction and the valence bands touch. The
low-energy effective Hamiltonian of these WSMs around these Weyl
points is given by $H= \pm \hbar v_F {\vec \tau} \cdot {\vec k}$,
where $\vec k= (k_x, k_y, k_z)$ is the wave vector, $v_F$ is the
velocity of the electrons near the Weyl point which depends on
material parameters, and $\vec \tau$ denotes Pauli matrices. These
Weyl nodes occur in pairs and are protected due to either
time-reversal or inversion symmetry breaking \cite{weylrev}. Such
isotropic Weyl nodes are characterized by a topological winding
number which takes values $\pm 1$ depending on the chirality of the
electrons around the node. The electron around such nodes display
spin momentum locking; this property along with the linear
dispersion $E_{\vec k} = \pm \hbar v_F |\vec k|$ and a non-zero
topological winding number distinguishes WSMs from ordinary metals.
This distinction is manifested in several unconventional features
associated with transport, magneto-transport and edge physics of
these materials \cite{weylrev,
transport1,transport2,transport3,edge1}.

More recently, materials with Weyl points having anisotropic
dispersion in two transverse direction (chosen to be $k_x$ and $k_y$
in this work) has been discovered \cite{msm1}. Such materials are
termed as multi-Weyl semimetals (MSMs) since their anisotropic
dispersion occurs due to merger of two or more Weyl nodes with same
chirality. Such a merger is found to be topologically protected by
point group symmetries (such as $C_4$ and $C_6$ rotational
symmetries) \cite{msm2}. The low energy dispersion of the electrons
in MSMs remain linear in the symmetry direction (chosen to be $k_z$
in this work) but vanishes as $k^n$ (where $k=\sqrt{k_x^2+k_y^2}$)
with $n>1$ in the transverse directions: $E(k_z=0, k) \sim k^n$
[\onlinecite{msm1,msm2,msm3}]. The topological winding number of
these anisotropic Weyl points is given by an integer $n$ with $n \le
3$ [\onlinecite{msm2}]. The presence of a winding number different
from unity modifies the helicity properties and the density of
states of the electrons in these materials \cite{msmheli}. Several
other signatures of $n \ne 1$ show up in optical and transport
quantities such as longitudinal optical conductivity, anomalous Hall
conductivity, collective modes, and magnetoresistance
\cite{msmoptical,msmcollective, msmmagneto}.

It is well known that transport measurement across junctions of
topological materials provides access to their topological
properties and unravels several unconventional features that have no
analog in standard metals \cite{geim1, beenakker1,
ks1,ks2,beenakker2,weyl1}. In 2D topological materials such as
graphene, the tunneling conductance $G$ across graphene normal
metal-barrier-normal metal (NBN) junctions, display oscillatory
behavior and a transmission resonance as a function of the barrier
potential \cite{geim1}. Similar behavior is also seen in subgap
tunneling conductance of graphene NBS junctions, where
superconductivity is induced in graphene via a proximate $s$-wave
superconductor \cite{ks1}. Such an oscillatory behavior and the
transmission resonance phenomenon turns out to be a signature of the
Dirac quasiparticles in graphene; they do not occur in standard
metals with Schrodinger quasiparticles in the regime where the
incident energy of a quasiparticle is small compared to the barrier
height. Similar behavior is also seen for quasiparticles on the
surface of a topological insulator \cite{ks2}. More recently
tunneling conductance across NBN and NBS junctions of WSMs have also
been studied \cite{weyl1, weyl2,weyl3, weyl4}. In particular, it was
found that the NBS junctions of time-reversal symmetric Weyl
semimetals may host a universal zero-bias conductance value of
$e^2/h$. In addition, the subgap tunneling conductance is found to
oscillate as a function of the barrier strength as expected for
standard Dirac materials \cite{weyl3}.

In this work, we study the tunneling conductance across NBN and NBS
junctions between either a WSM ($n=1$) and a MSM ($n \ne 1$) or two
MSMs with $n_1 \ne n_2$ separated by a barrier of width $d$ and a
potential $U_0$. Such junctions differ from their previously studied
WSM counterparts in the sense that the topological winding number of
the system changes across these junctions. The main results obtained
from our study are as follows. First, we show that the tunneling
conductance $G$ of these junctions becomes independent of the
barrier potential in the thin barrier limit where $U_0 \to \infty$
and $d \to 0$ with $\chi = U_0 d/(\hbar v_F)$ being held fixed. We
note that this behavior is in contrast to that found in junctions of
both ordinary Schrodinger metals (where $G$ is a monotonically
decaying function of $\chi$) and Dirac or WSM materials (where $G$
oscillates with $\chi$). We demonstrate that this independence is a
consequence of difference of winding numbers between the WSM and MSM
(or two MSMs) on two sides of the junction. Second, we find that the
subgap tunneling conductance of the NBS junction depends crucially
on the topological winding numbers. It vanishes if superconductivity
is induced on the MSM with higher topological winding number; in
contrast, it is finite when superconductivity is induced on the WSM
or MSM with lower topological winding number. Third, we analyze the
fate of the tunneling conductance $G$ for these junctions away from
the thin barrier limit. We find that they display weak oscillatory
dependence on the barrier potential $U_0$ for finite barrier
thickness $d$; the amplitude (period) of these oscillations
decreases (increases) with $d$ for any finite $U_0$. For large
$U_0$, $G$ becomes independent of $U_0$ leading to the thin barrier
result.  Finally, we discuss experiments which can test our theory.

The plan of the rest of the paper is as follows. In Sec.\ \ref{nbn},
we analyze the transport in NBN junctions between a WSM and a MSM or
two MSMs with different winding numbers. This is followed by a
similar analysis for NBS junctions in Sec.\ \ref{nbs}. Finally, we
discuss our main results, point out relevant experiments which may
test our theory, and conclude in Sec.\ \ref{diss}. We detail some of
the calculation regarding inter-node scattering in the Appendix.

\begin{figure}
{\includegraphics[width=0.98\hsize]{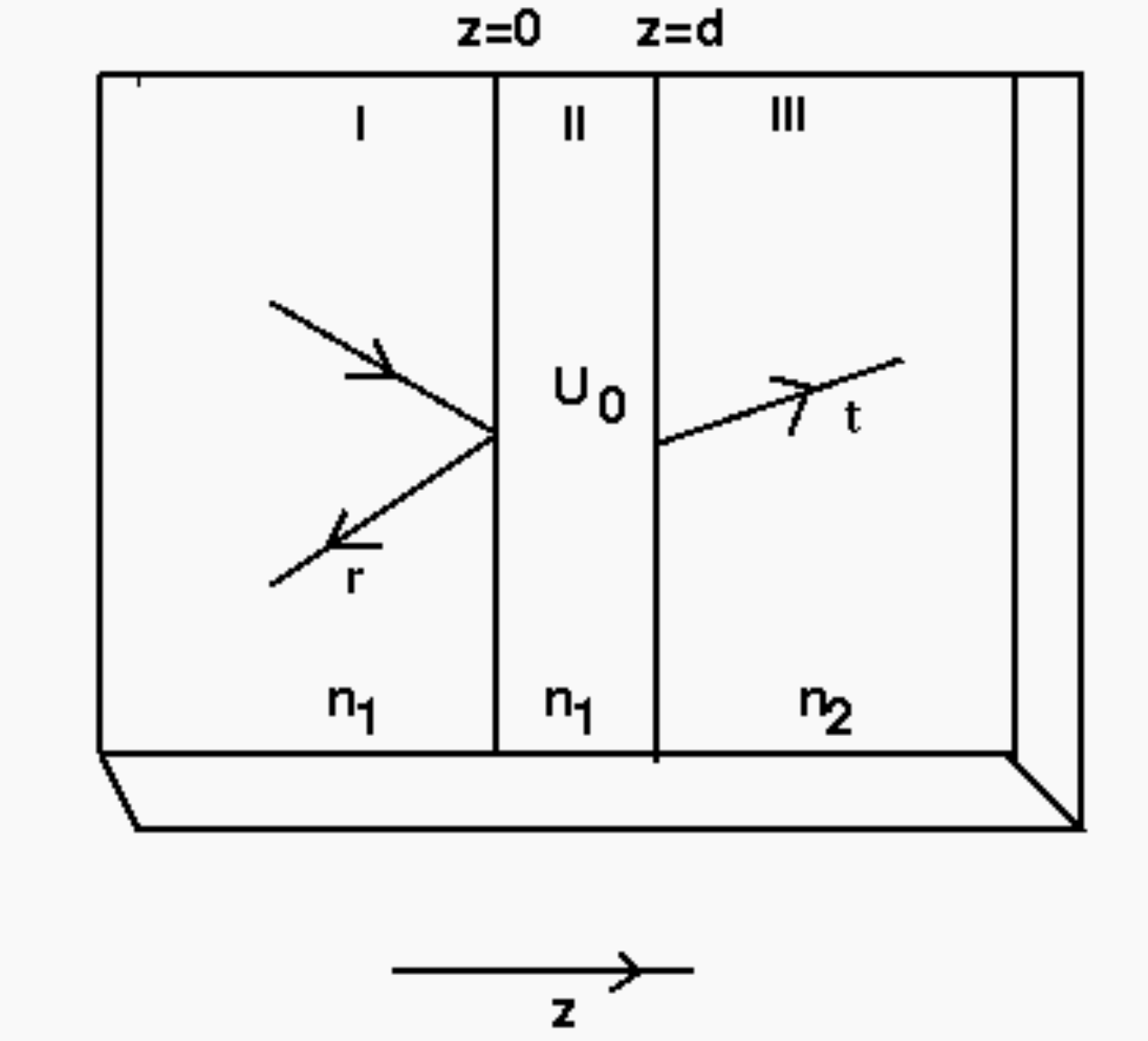}} \caption{ A schemetaic
representation of a junction between two MSMs or a WSM and a MSM
(characterized by topological winding numbers $n_1$ and $n_2$ as
shown) in regions I and III separated by a barrier in region II. The
barrier region II extending from $z=0$ to $z=d$ constitutes the same
material as in region I but has an additional potential $U_0$.  The
reflection  and transmission amplitudes, $r$ and $t$, for an
electron approaching the barrier from region I is shown
schematically. \label{fig1}}
\end{figure}

\section{NBN junctions}
\label{nbn}

In this section we shall derive the conductance of a NBN junction
between a WSM and a MSM or two MSMs with different winding numbers.
The geometry of the setup is sketched in Fig.\ \ref{fig1}. The
Hamiltonian of the system is given by
\begin{eqnarray}
H &=& H_1 \theta(d-z) + U_0 \theta(d-z)\theta(z) + H_2 \theta(z-d)
\label{hamnbn}
\end{eqnarray}
where $\theta(z)$ is the Heaviside step function. The Hamiltonians
$H_1$ and $H_2$ are given by
\begin{eqnarray}
H_1 &=& E_0 (- i \partial_z  \tau_z  + \epsilon_0 k^{n_1}
\nonumber\\
&& \times \left[\cos(n_1 \phi_k) \tau_x + \tau_y \sin(n_1 \phi_k)
\right] )
\nonumber\\
H_2 &=& \eta^{-1} E_0 (- i \partial_z  \tau_z  + \epsilon'_0
k^{n_2} \nonumber\\
&& \times \left[\cos(n_2 \phi_k) \tau_x + \tau_y \sin(n_2 \phi_k)
\right] ), \label{weylham}
\end{eqnarray}
where $n_1$ and $n_2$ are the topological winding numbers in regions
I and II as shown in Fig.\ \ref{fig1}, $\phi_k = \arctan(k_y/k_x)$,
and $E_0= \hbar v_F k_0$ is the energy scale in which all energies
are measured. In the rest of this work, we shall take this energy
scale to be upper cutoff up to which the low-energy continuum
Hamiltonians (Eq.\ \ref{weylham}) hold. Here $v_F$ and $v'_F =v_F/
\eta $ are the Fermi velocities for electrons in region I and III,
$k_0$ is the momentum scale chosen to make all momenta
dimensionless, and $\epsilon_0$ and $\epsilon'_0$ are material
specific constants whose precise numerical value is not going to
alter our main results. We shall further choose a common chemical
potential $\mu_N$ across the junction. We note that the analysis we
carry out holds even if the chemical potentials in regions I and III
are different; we nevertheless choose them to be the same to reduce
the number of parameters in the theory. Moreover it is always
possible to align the chemical potentials in regions I and III by
applying a voltage across one of them. In what follows, we shall
apply a voltage $V$ across the junction and compute $G$ as a
function of $V$.

To compute $G$, we first consider the electron wavefunction in
region I. A straightforward calculation shows that the wavefunction
for right(R) and left(L) moving electrons in region I in the
presence of an applied voltage $eV$ is given by \cite{weyl3}
\begin{eqnarray}
\psi_{1 e R} &=& e^{ i(k_x x + k_y y + k_{z1} z)} e^{-i \tau_z n_1
\phi_k/2}( \cos(\theta_1), \sin(\theta_1))^T
\nonumber\\
\psi_{1 e L} &=& e^{ i(k_x x + k_y y - k_{z1} z)} e^{-i \tau_z n_1
\phi_k/2} ( \sin(\theta_1), \cos(\theta_1))^T \nonumber\\
\label{ewav1}
\end{eqnarray}
where $2\theta_1 = \arcsin(\epsilon_0 k^{n_1}/|eV+\mu_N|)$, $k=
\sqrt{k_x^2+k_y^2}$, $k_{z1} = {\rm Sgn}(eV+ \mu_N + \epsilon_0
k^{n_1}) \sqrt{ (eV+\mu_N)^2-\epsilon_0^2 k^{2n_1}}$ and we have
measured all energies (wavevectors) in units of $E_0=\hbar v_F k_0$
($k_0$). We note that here and in the rest of this work, the
temporal dependence of the wavefunctions [{\it i.e} $\exp[-i
(eV)t/\hbar]$ factor] has not been explicitly mentioned for clarity.

The wavefunction in region I can be written in terms of $\psi_{eR}$
and $\psi_{eL}$ as
\begin{eqnarray}
\psi_1 &=& \psi_{1eR} + r \psi_{1eL}  \label{wavreg1}
\end{eqnarray}
where $r$ is the amplitude of reflection from the barrier. We note
here that $\psi_{1 e R(L)} \sim e^{-i \tau_z n_1 \phi_k/2}$ leading
to $\psi_1 \sim e^{-i \tau_z n_1 \phi_k/2}$; thus the azimuthal
angle dependence of the wavefunction in region I can be interpreted
as a spin rotation by an angle of $n_1 \phi_k$ about the $\hat z$
axis.

In region II, the electrons see an additional applied potential
$U_0$. The right and the left moving electron wavefunction in this
regime can be written as
\begin{eqnarray}
\psi_{2 e R} &=& e^{ i(k_x x + k_y y + k_{z2} z)} e^{-i \tau_z n_1
\phi_k/2} (\cos(\theta_2), \sin(\theta_2))^T
\nonumber\\
\psi_{2 e L} &=& e^{ i(k_x x + k_y y - k_{z2} z)} e^{-i \tau_z n_1
\phi_k/2} (\sin(\theta_2), \cos(\theta_2))^T \nonumber\\
\label{ewav1}
\end{eqnarray}
where $k_{z2} = {\rm Sgn}(eV +\mu_N -U_0 + \epsilon_0 k^{n_1})
\sqrt{ (eV+\mu_N - U_0)^2-\epsilon_0^2 k^{2n_1}}$ and $ 2 \theta_2 =
\arcsin (\epsilon_0 k^{n1}/|eV+\mu_N-U_0|)$. We note that $\theta_2
\to 0$ when $U_0 \to \infty $. Thus the wavefunction in region II
can be written as
\begin{eqnarray}
\psi_{II} &=&  p \psi_{2eR} + q \psi_{2 eL}  \label{wavreg2}
\end{eqnarray}
where $p$ and $q$ denotes amplitudes of right and left moving
electrons in region II. We note that $\theta_2$ and $\theta_1$ are
related by
\begin{eqnarray}
\theta_2 =\frac{1}{2}\arcsin[ \sin(2 \theta_1)/|1-U_0/(eV+\mu_N)|]
\label{thetarel1}
\end{eqnarray}

In region III, the right moving electrons have a wavefunction given
by
\begin{eqnarray}
\psi_{3 e R} &=& e^{ i(k_x x + k_y y + k_{z3} z)} e^{-i \tau_z n_2
\phi_k/2} ( \cos(\theta_3), \sin(\theta_3))^T \nonumber\\
\label{ewav3}
\end{eqnarray}
where $2\theta_3 = \arcsin(\epsilon'_0 k^{n_2}/|\eta(eV+\mu_N)|)$,
$k_{z3} = {\rm Sgn}[(eV + \mu_N)\eta + \epsilon'_0 k^{n_2}] \sqrt{
(eV+\mu_N)^2 \eta^2 -\epsilon_0^{'2} k^{2n_2}}$, and $\eta$ is the
measure of the Fermi velocity mismatch across the junction. We note
that $\theta_1$ and $\theta_3$, for any given voltage $eV$, are
related by
\begin{eqnarray}
\sin(2\theta_3)  &=& |eV+ \mu_N|^{n_2/n_1 -1} \sin[2
\theta_1]^{n_2/n_1} \frac{\epsilon'_0}{\eta \epsilon_0^{n_2/n_1}}
\label{thetarel2}
\end{eqnarray}
The wavefunction in region III is thus given by
\begin{eqnarray}
\psi_{III} = t \psi_{3eR} \label{wavreg3}
\end{eqnarray}
where $t$ is the transmission amplitude across the junction.  We
note that $\psi_3 \sim e^{-i \tau_z n_2 \phi_k/2}$.

To obtain the reflection and transmission amplitude across the
barrier, we match the wavefunctions at $z=0$ and $z=d$, where $d
\equiv dk_0$ constitutes the width of the barrier in units of
$k_0^{-1}$. This requires $\psi_I(z=0)=\psi_{II}(z=0)$ and
$\psi_{II}(z=d)=\psi_{III}(z=d)$ and leads to
\begin{widetext}
\begin{eqnarray}
&&\cos(\theta_1) + r \sin(\theta_1) = p
\cos(\theta_2) + q \sin(\theta_2)  \nonumber\\
&&\sin(\theta_1) + r \cos(\theta_1) = p \sin(\theta_2)
+ q \cos(\theta_2) \nonumber\\
&&p \cos(\theta_2)e^{i k_{z2} d} + q \sin(\theta_2) e^{-i k_{z2} d} = t \cos(\theta_3) e^{i (k_{z3} d -(n_2-n_1)\phi_k/2)} \nonumber\\
&&p \sin(\theta_2)e^{i k_{z2} d } + q \cos(\theta_2) e^{-i k_{z2} d}
= t \sin(\theta_3) e^{i (k_{z3} d +(n_2-n_1) \phi_k/2)}
\label{wavmatchnbn}
\end{eqnarray}
\end{widetext}
Solving for $r$ from these equations one obtains $r= {\mathcal
N}/{\mathcal D}$ where
\begin{eqnarray}
&&{\mathcal N} = \cos(\theta_3) \big[ \sin(\theta_1)
+\sin(\theta_1-2\theta_2) + 2 e^{2ik_{z2}d} \nonumber\\
&& \times \sin(\theta_2)\cos(\theta_1+\theta_2) \big]
+\sin(\theta_3) e^{i(n_2-n_1)\phi_k}  \big[ \cos(\theta_1) \nonumber\\
&& -\cos(\theta_1-2\theta_2) - 2e^{2ik_{z2} d} \cos(\theta_2)
\cos(\theta_1+\theta_2) \big]
\nonumber\\
&&{\mathcal D}= 2\Big[ e^{i(n_2-n_1)\phi_k} \sin(\theta_3) \big[
e^{2i d k_{z2}} \cos(\theta_2) \sin(\theta_1-\theta_2) \nonumber\\
&& + \sin(\theta_2) \cos(\theta_1+\theta_2) \big] - \cos(\theta_3)
\big[ \cos(\theta_2) \cos(\theta_1+\theta_2) \nonumber\\
&& + e^{2 id k_{z2}} \sin(\theta_2) \sin(\theta_1-\theta_2) \big]
\Big]\label{numden}
\end{eqnarray}
The expression of the transmission and hence the conductance can be
obtained using Eq.\ \ref{numden} as $T= 1-|r|^2$ and
\begin{eqnarray}
G &=& G_0 \int_{0}^{k_{\rm max}} \frac{k dk}{2\pi} \int_0^{ 2 \pi}
\frac{d \phi_k}{2\pi} \, T
\nonumber\\
G_0 &=& \frac{n_0 e^2}{h N_1}, \quad N_1 = \int_{0}^{1} \frac{k
dk}{2\pi} \int_0^{ 2 \pi} \frac{d \phi_k}{2\pi}  = \frac{1}{4 \pi}.
\label{condrel1}
\end{eqnarray}
Here $N_1 \equiv N_1 k_0^2$ denote the total number of transverse
modes around a Weyl node up to the cutoff $k_0$ for which the
continuum Weyl model used here holds, $n_0$ is the total number of
Weyl nodes in the Brillouin zone each of which provides independent
contribution to $G$, and $k_{\rm max} = {\rm
Min}[[(eV+\mu_N)/\epsilon_0]^{1/n_1},
[\eta(eV+\mu_N)/\epsilon'_0]^{1/n_2}]$ is the largest momentum
channel participating in current transport across the junction. Note
that $k_{\rm max}$ is determined by the condition that both
$\theta_1 = \arcsin[\epsilon_0 k^{n_1}/|eV+\mu_N|]/2$ and $\theta_3
= \arcsin[\epsilon'_0 k^{n_2}/(\eta|eV+\mu_N|)]/2$ must be real for
a particular momentum channel to conduct.

\begin{figure}
{\includegraphics[width=0.8\hsize]{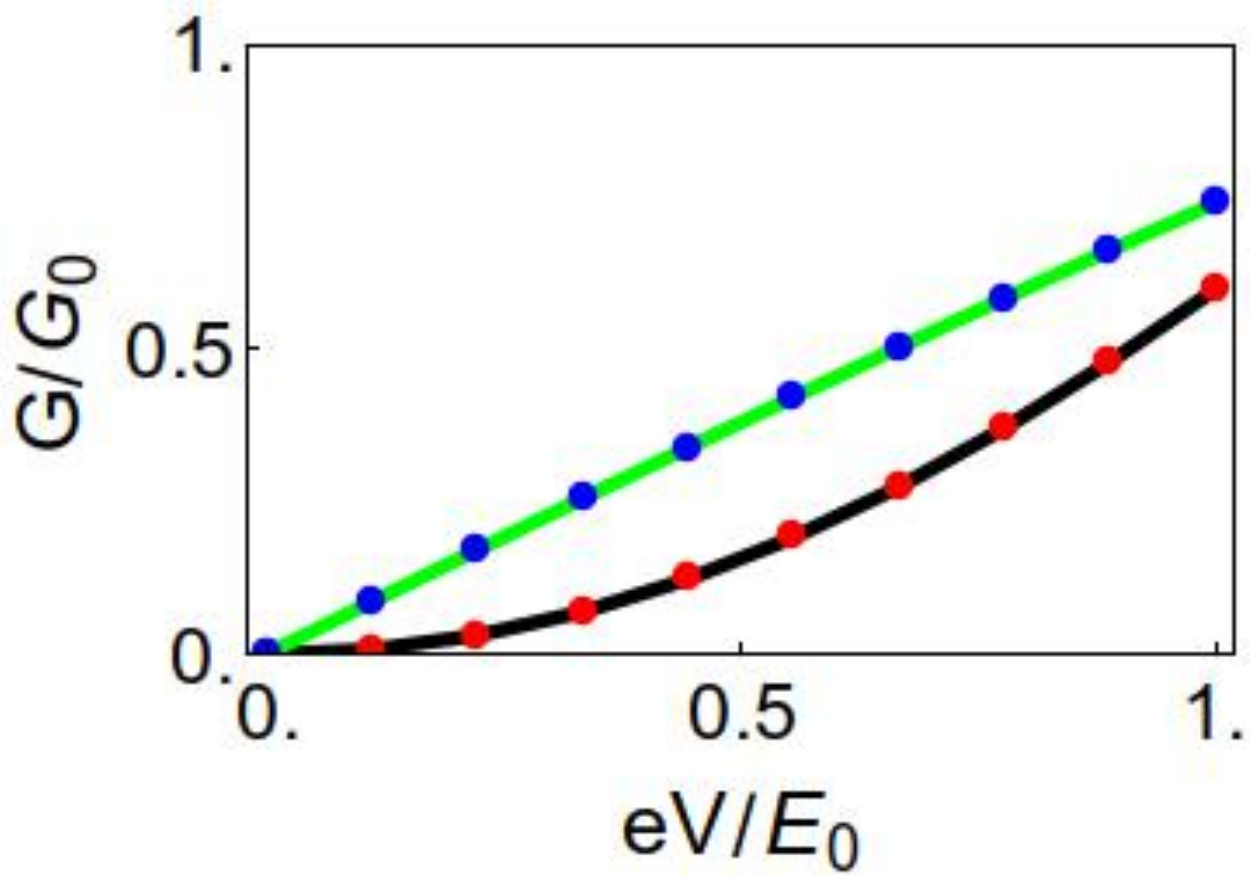}} \caption{ (a) Plot of
$G/G_0$ as a function of $eV$ (in units of $E_0$) for $\delta n=\pm
1$. The black solid line (red dots) corresponds to $n_1=1$, $n_2=2$,
and $\chi=0 (\pi/4)$ while the green solid line (blue dots)
corresponds to $n_1=2$, $n_2=3$ and $\chi=0(\pi/4)$. For all plots,
$\epsilon_0=\epsilon'_0=\eta=1$ and $\mu_N=0$. See text for details.
\label{fig2}}
\end{figure}

Next, we note that in contrast to junctions between WSMs or two
similar MSMs with $n_1=n_2$, $|r|^2$, and hence $T$ possess
non-trivial $\phi_k$ dependence for the present junctions where $n_1
\ne n_2$. To understand this phenomenon better, we now move to the
thin barrier limit. In this limit, it is easy to see that
$\theta_2,\, k_{z3}d \to 0$, and $k_{z2}d \to -\chi$. The boundary
conditions can then be written as
\begin{eqnarray}
\psi_I(z=0^-) = e^{i \tau_3 \chi} \psi_{III} (z=0^{+})
\label{bcmatrix}
\end{eqnarray}
We note that this implies that the dimensionless barrier potential
induces a rotation by $-2 \chi$ in spin space about the $z$ axis.
For $n_1=n_2$, this leads to oscillatory dependence of the
conductance on $\chi$. In contrast, for $n_1 \ne n_2$, since
$\psi_{I[III]} \sim \exp[-i \tau_z n_1[n_2] \phi_k/2]$, the rotation
induced by the barrier can be offset by changing $\phi_k \to \phi_k
+\delta \phi$, where $\delta \phi = 2\chi/(n_2-n_1)$. Thus the
junction conductance, which involves a sum over all azimuthal
angles, is expected to become barrier independent in the thin
barrier limit.

To verify this expectation, we first substitute $\theta_2,\, k_{z3}d
\to 0$, and $k_{z2}d \to \chi$ in Eq.\ \ref{numden} and obtain,
after a few lines of algebra,
\begin{eqnarray}
T_{\rm tb} &=& \frac{{\mathcal A}}{{\mathcal B} -{\mathcal C}
\cos[(n_1-n_2)\phi_k +2\chi]} \nonumber \\
{\mathcal A} &=& \cos(2 \theta_1) \cos(2 \theta_3), \quad
{\mathcal C} = \sin(2 \theta_1)\sin(2 \theta_3)/2 \nonumber\\
{\mathcal B} &=& \sin^2(\theta_1) \sin^2(\theta_3) +
\cos^2(\theta_1) \cos^2(\theta_3) \label{tamptb}
\end{eqnarray}
From Eq.\ \ref{tamptb}, we find that in the presence of a change in
winding number across region I and III ($n_1 \ne n_2$), $\chi$
appears as a phase shift to the azimuthal angle $\phi_k$. Since
${\mathcal A}$, ${\mathcal B}$, and ${\mathcal C}$ are independent
of $\phi_k$, the integration over $\phi_k$ in Eq.\ \ref{condrel1} is
straightforward and yields
\begin{eqnarray}
\int_0^{ 2 \pi} \frac{d \phi_k}{2\pi} T_{tb} &=&  T_1, \quad G = G_0
\int_0^{k_{\rm max} } \frac{k dk}{2\pi}\, T_1 \nonumber\\
T_1 &=& \frac{2 \cos(2 \theta_1) \cos(2 \theta_3)}{|\cos(2 \theta_1)
+ \cos(2\theta_3)|} \label{tbcond}
\end{eqnarray}
We therefore find that $G$ becomes independent of $\chi$ in the thin
barrier limit according to our earlier expectation. This
independence is a direct consequence of $\phi_k$ dependence of $T$
which happens for $n_1\ne n_2$. Thus such a barrier independence of
$G$ requires a change in the topological winding number across the
junction; consequently, this effect would not show up in junctions
between WSMs or MSMs with $n_1=n_2$. We would like to point out that
this phenomenon can only occur in $d>2$ where there are more than
one transverse directions; thus it does not have an analogue in 2D
topological materials.

\begin{figure}
{\includegraphics[width=0.48\hsize]{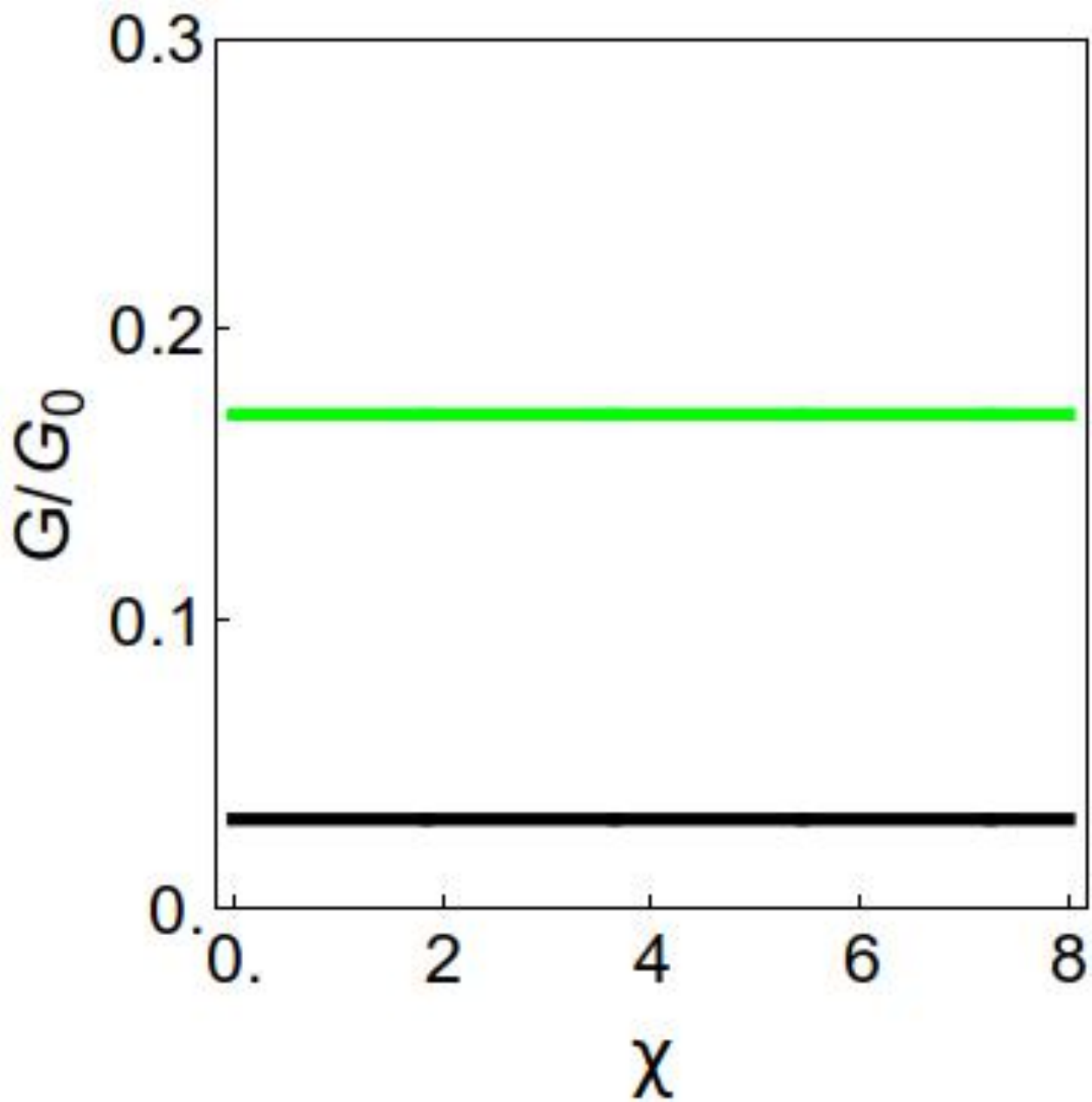}}
{\includegraphics[width=0.48 \hsize]{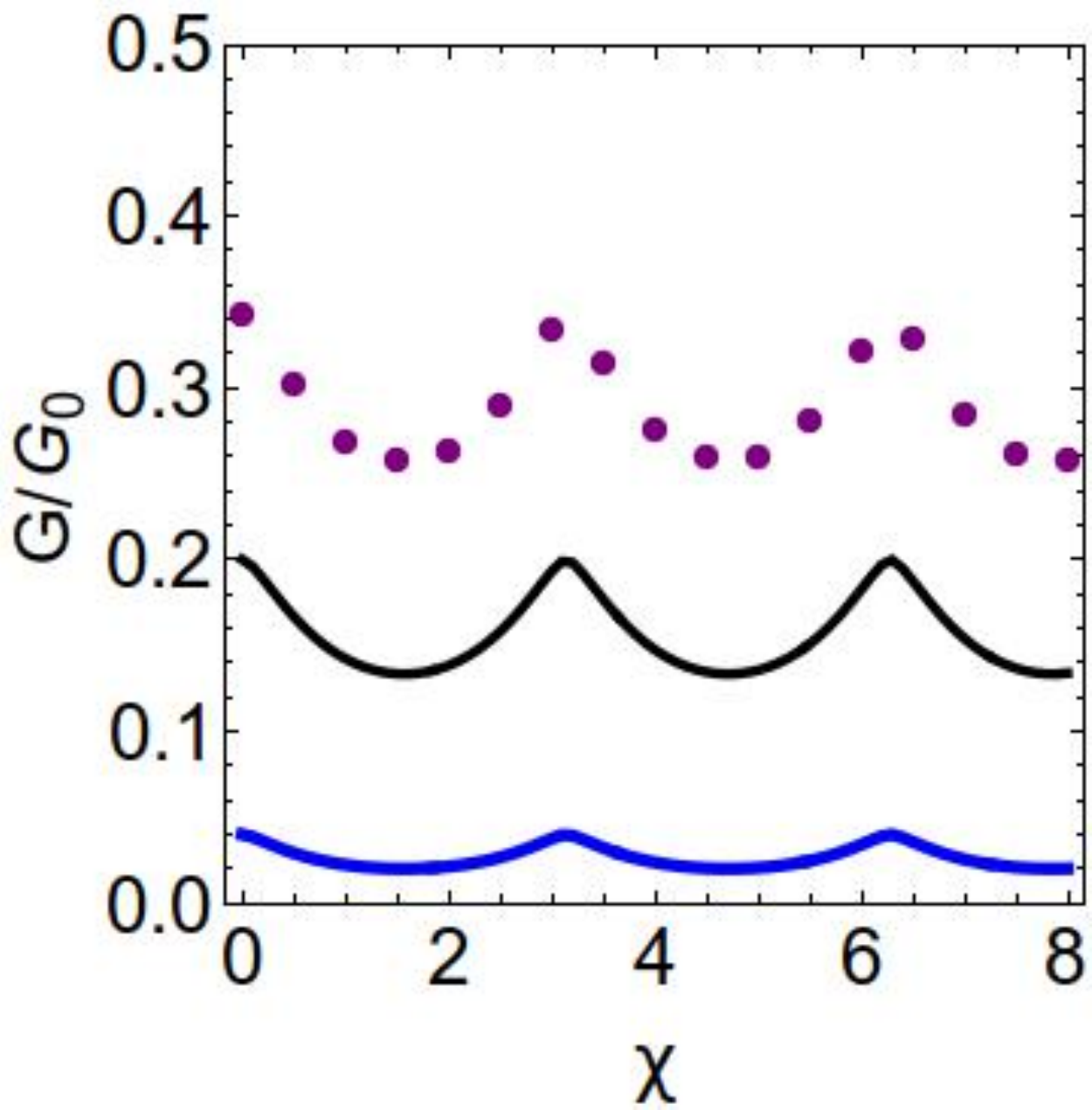}} \caption{ (a) Plot of
$G/G_0$ as a function of $\chi$ for $eV/E_0=0.2$. The black (green)
solid lines correspond to $n_1=1(2)$ and $n_2=2(3)$. (b) Plot of
$G/G_0$ as a function of $\chi$ for $n_1=n_2=n_0$ and $eV/E_0=0.2$.
The blue solid, the black dashed, and the magenta dotted lines
correspond to $n_0=1$, $2$ and $3$ respectively. All other
parameters are same as in Fig.\ \ref{fig2}. See text for details.
\label{fig3}}
\end{figure}

Next, we provide numerical support to our finding. To this end, we
first obtain $G/G_0$ by numerically integrating $T_{tb}$ over $k$
and $\phi_k$. For all numerical plots we shall choose
$\eta=\epsilon'_0=\epsilon_0=1$; we have checked that the numerical
values of these quantities do not alter qualitative nature of the
results presented. The corresponding results are shown in Fig.\
\ref{fig2} and \ref{fig3}. In Fig.\ \ref{fig2}, we show the
variation of $G/G_0$ as a function of the applied voltage $eV/E_0$
in the thin barrier limit for $n_1-n_2 =-1$ with $n_1=1$ and $n_1=2$
and for two representative values of $\chi=0, \pi/4$. We have
checked that the behavior of $G$ is identical for $n_1-n_2=1$ and
qualitatively similar for $n_1-n_2=\pm 2$ for same $n_1$.  The
different behavior of $G$ as a function of $eV/E_0$ for $n_1=1$ and
$n_1=2$ can be understood as follows. We note from Eq.\ \ref{tamptb}
and \ref{tbcond} that for small $eV$, $\theta_3 \ll \theta_1$ (Eq.\
\ref{thetarel2}). Consequently one may approximate
\begin{eqnarray}
T_1 \simeq \frac{2 \cos(2 \theta_1)}{|1+\cos(2 \theta_1)|} = \frac{2
\sqrt{1 - k^{2n_1}/(eV+\mu_0)^2}}{1+  \sqrt{1 -
k^{2n_1}/(eV+\mu_0)^2}} \label{t1approx}
\end{eqnarray}
The integral $T_1$ over $k$ can then be analytically performed and
leads to $ G/G_0 \sim k_{\rm max}^2 = c |eV+\mu_0|^{2/n_1}$, where
$c$ is a constant. Thus $G/G_0$ is a parabolic (linear) function of
the applied voltage for $n_1=1(2)$ and $\mu_N=0$. An exactly similar
behavior emerges when $n_1 > n_2$ since $T_1$ is symmetric under the
interchange of $\theta_3$ and $\theta_1$. Note that for finite
$\mu_N/E_0 <1$ and $eV \ll \mu_N$, $G/G_0$ will always vary linearly
with $eV$.

For both $n_1=1$ and $n_1=2$, from Fig.\ \ref{fig2}, we find that
$G/G_0$ is independent of $\chi$. This independence can be more
directly seen from Fig.\ \ref{fig3}(a). We also note that such a
barrier independence is absent if $n_1=n_2$; this is easily seen
from Fig.\ \ref{fig3}(b), where $G$ oscillates with $\chi$ for a
junction between two WSMs ($n_1=n_2=1$) or MSMs ($n_1=n_2=2,3$). We
note that these numerical results confirm our earlier analytical
expectation that the $\chi$ independence of $G$ is a consequence of
the change in topological winding number across the junction.

\begin{figure}
{\includegraphics[width=0.8 \hsize]{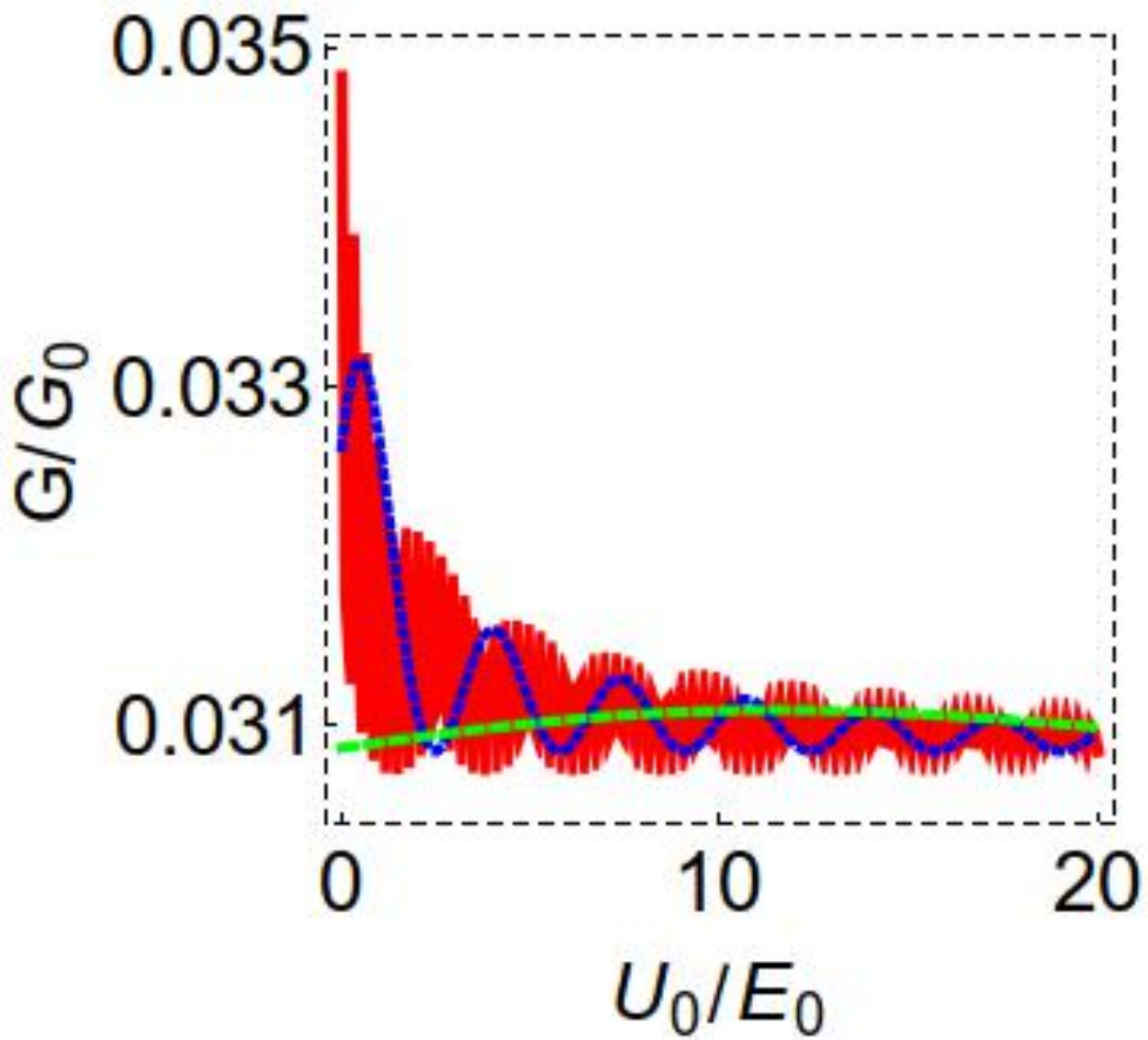}} \caption{ Plot of $G/G_0$
as a function of $U_0/E_0$ for $dk_0=10$ (red solid line), $1$ (blue
dotted line) and $0.1$ (green dashed line). For all plots $eV/E_0=
0.2$, $n_1=1$, and $n_2=2$. All other parameters are same as in
Fig.\ \ref{fig2}. \label{fig4}}
\end{figure}

Next, we investigate the fate of $G$ as a function of $U_0$ away
from the thin barrier limit for several representative values of
$d$. To this end, we numerically compute $T=1-|r|^2$ from Eq.\
\ref{numden} and use Eq.\ \ref{condrel1} to obtain $G$. Fig.\
\ref{fig4} shows a plot of $G/G_0$ as a function of $U_0$ for
several representative values of $d$. We note that $G/G_0$ has small
oscillatory dependence on $U_0$; the amplitude of these oscillations
decay as $U_0$ is increased and $G/G_0$ becomes independent of $U_0$
for large $U_0$. This is consistent with our earlier results in the
thin barrier limit.

Before ending this section, we observe a few points regarding our
analysis. First, we have carried out this analysis for ballistic
junctions. The justification for such an analysis is two fold.
First, it is well-known that Weyl semimetals, in the presence of
weak disorder, hosts a quasi-ballistic regime \cite{disorder1}. In
this regime, we expect the computations carried out in the ballistic
regime to be qualitatively correct as demonstrated earlier. For this
one needs the typical mean-free path to be larger than the barrier
region \cite{disorder2}. Moreover we note that a scalar disorder
potential in Weyl semimetals can not scatter between states with
different $\phi_{\vec k}$. This can be seen through a direct
calculation using Gaussian disorder potential \cite{disorder1,
disorder2}; however, one can also understand this by noting that
such a scattering would correspond to a spin rotation about $\hat z$
which a scalar (spin-independent) disorder potential can not
achieve. Thus the $\phi_{\vec k}$ dependence of the wavefunction
which is central to the barrier independence discussed above is
expected to be robust in the weak disorder regime. Second, we have
neglected inter-node scattering. We note that as long as these Weyl
nodes occur at different transverse momentum, the barrier can not
lead to such inter-node scattering (since a barrier potential
conserve transverse momenta on scattering). However, if the nodes
occur at same transverse momentum such inter-node scattering can
occur; we analyze this situation taking a simple model with two Weyl
nodes at $\vec k= (0,0 \pm K_0)$ in the Appendix. We show that long
as $K_0 \gg d/a^2, eV/\hbar v_F, U_0/\hbar v_F$, such inter-node
scattering is suppressed. Thus our analysis holds for a wide range
of parameters which we chart out in the Appendix.

\section{NBS junctions}
\label{nbs}

\begin{figure}
{\includegraphics[width=0.9\hsize]{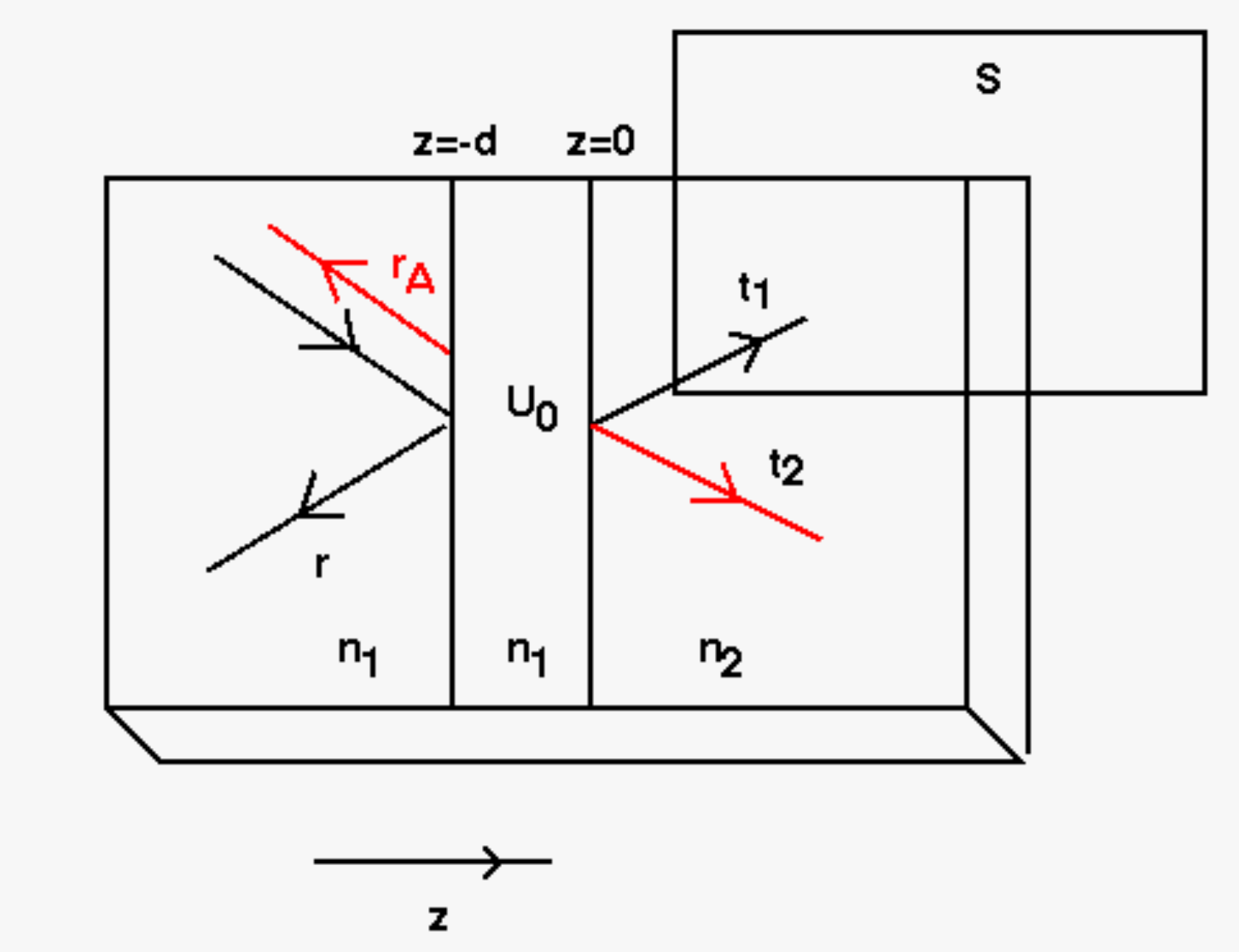}} \caption{ A schemetaic
representation of a NBS junction between two MSMs or a WSM and a MSM
(characterized by topological winding numbers $n_1$ and $n_2$ as
shown) in regions I and III separated by a barrier in region II
($-d\le z\le 0$). The barrier region II constitutes the same
material as in region I but has an additional potential $U_0$.
Superconductivity is induced in region III via a proximate $s$-wave
superconductor. The reflection, Andreev reflections  and
transmission amplitudes, $r$, $r_A$, $t_1$ and $2$, for an electron
approaching the barrier from region I, is shown schematically.
\label{fig5}}
\end{figure}

In this section, we study the transport through a NBS junction
between a WSM and a MSM or two MSMs with different topological
winding numbers. Throughout this section we shall work in the regime
where the chemical potential $\mu_S$ in the superconduction region
is large compared to the applied voltages but is small compared to
$E_0$. We note that the chemical potential $\mu_N$ shall be kept
arbitrary. The schematic representation of such a junction is given
by Fig.\ \ref{fig5}. The MSM (or WSM) in region III has a proximate
$s-$wave superconductor. In this section, we shall consider,
following Ref.\ \onlinecite{weyl3}, the case where the induced
superconductivity is $s-$wave and the Cooper pairing connects two
isotropic or anisotropic Weyl nodes with same chirality. The basis
of this choice is the observation made in Ref.\ \onlinecite{weyl3}
that the inter-orbital superconduction pairing between nodes of
opposite chirality is suppressed at low energy. We note that
necessitates the presence of at least four Weyl or multi-Weyl nodes
in region III. With this model of induced superconductivity the
Hamiltonian in region III is given by a $4 \times 4$ matrix
\begin{eqnarray}
H_2^s &=&  (H_2-\mu_S) \otimes \sigma_3 + E_0 \eta^{-1} \Delta_0 I
\otimes \sigma_1, \label{supham3}
\end{eqnarray}
where $\sigma_i$ for $i=1,2,3$ denote Pauli matrices in
particle-hole space, $\mu_S \gg eV, \Delta_0$ is the chemical
potential, and $H_2$, given by Eq.\ \ref{weylham}, may represent a
WSM or a MSM depending on the value of $n_2$. Here we shall choose
the phase of the superconduction condensate to be zero without any
loss of generality. The basic excitations of $H_2^s$ are Bogoliubov
quasiparticles and quasiholes. The wavefunction of such right-moving
quasiparticles, which would be necessary for our computation, are
given by
\begin{widetext}
\begin{eqnarray}
\psi^s_{3e} &=&  e^{-i \tau_z n_2 \phi_k/2} \left(e^{i \mu_0} \cos
\theta_3^s, e^{i \mu_0 }\sin \theta_3^s, \cos \theta_3^s,
\sin\theta_3^s \right)^T e^{i(k_x x + k_y y + k^s_{3z1} z)}/\sqrt{2} \nonumber\\
\psi^s_{3h} &=& e^{-i \tau_z n_2 \phi_k/2} \left( \cos
\theta_3^{'s}, \sin \theta_3^{'s}, e^{i\mu_0} \cos \theta_3^{'s},
e^{i \mu_0} \sin\theta_3^{'s} \right)^T   e^{i(k_x x + k_y y +
k^{s}_{3z2} z)}/\sqrt{2} \label{squasi1}
\end{eqnarray}
\end{widetext}
where $\tan(2 \theta^s_{3}[\theta_3^{'s}])= \epsilon'_0
k^{n_2}/k^s_{3z1[2]}$. In Eq.\ \ref{squasi1}, $k^s_{3z1[2]}$
correspond to electron-[hole-]like quasiparticles and are given by
(for $\mu_S \gg eV, \Delta_0$)
\begin{eqnarray}
k^s_{3z1[2]} &=& +[-] \sqrt{(\mu_S \pm i \zeta)^2 \eta^2-
(\epsilon'_0
k^{n_2})^2} \nonumber\\
\zeta &=& 1[i] \sqrt{\Delta_0^2- (eV)^2}, \quad {\rm for}\, eV \le
[\ge] \Delta_0 \nonumber\\
\cos \mu_0 [\cosh \mu_0] &=& eV/\Delta_0\quad {\rm for} \, eV \le
[\ge] \Delta_0. \label{squasi2}
\end{eqnarray}
where we have scaled all energy scales by $E_0$. We note that the
wavefunctions of the quasiparticles and quasiholes retain the
property $ \psi^s_{3e(h)} \sim \exp[-i \tau_z n_2 \phi_k/2]$.

The computation of tunneling conductance for such a junction follows
the standard BTK formalism \cite{btk1} applied to topological
materials \cite{beenakker1,ks1, weyl1}. To this end, we consider a
right moving electron in region I approaching the barrier. Upon
reflection (Andreev reflection) from the barrier, a left moving
electron (hole) propagates to the left. The wavefunctions of these
electron and holes are given by
\begin{widetext}
\begin{eqnarray}
\psi_{1eR} &=& e^{ i(k_x x + k_y y + k_{z1} z)}  e^{-i \tau_z n_1
\phi_k/2} ( \cos(\theta_1), \sin(\theta_1),0,0)^T
\nonumber\\
\psi_{1 e L} &=& e^{ i(k_x x + k_y y - k_{z1} z)}  e^{-i \tau_z n_1
\phi_k/2}  (\sin(\theta_1), \cos(\theta_1),0,0)^T \nonumber\\
\psi_{1hR} &=&  e^{ i(k_x x + k_y y + k'_{z1} z)}  e^{-i \tau_z n_1
\phi_k/2} (0,0, - \sin(\theta'_1),\cos(\theta'_1))^T
\nonumber\\
\psi_{1 h L} &=& e^{ i(k_x x + k_y y - k'_{z1} z)}  e^{-i \tau_z n_1
\phi_k/2} ( 0, 0,
\cos(\theta'_1), -\sin(\theta'_1))^T \nonumber\\
\label{ehwav1}
\end{eqnarray}
\end{widetext}
where $2\theta'_1= \arcsin[\epsilon_0 k^{n_1}/|eV-\mu_N|]$ and
$k'_{z1}= {\rm Sgn}(eV-\mu_N + \epsilon_0 k^{n_1})
\sqrt{(eV-\mu_N)^2- \epsilon_0^2 k^{2 n_1}}$. In region I, the
wavefunction can then be written as
\begin{eqnarray}
\psi_I^{\rm nbs} &=& \psi_{1eR} + r \psi_{1eL} + r_A \psi_{1hL}
\label{wavnbs1}
\end{eqnarray}
where $r$ and $r_A$ denotes amplitude or ordinary and Andreev
reflections respectively. We note that $\sin(\theta'_1) = -
\sin(\theta_1) |(eV+\mu_N)|/|(eV-\mu_N)|$ so that $\theta'_1 \to
-\theta_1$ for $\mu_N \gg eV$. Moreover, we find that $\psi_1 \sim
e^{-i \tau_z n_1 \phi_k/2}$; thus for both electrons and holes, one
can interpret $\phi_k$ dependence of the wavefunctions as a rotation
in spin space about the $z$ axis.

In region II, the wavefunctions of right/left moving electrons and
holes are given by Eq.\ \ref{ehwav1} with $k_{z1} \to k_{z2}$,
$\theta_1 \to \theta_2$, $k'_{z1} \to k'_{z2}$, and $\theta'_1 \to
\theta'_2$, where
\begin{eqnarray}
2\theta'_2 &=& \arctan[\epsilon_0 k^{n_1}/k'_{z2}] \nonumber\\
k'_{z2} &=& {\rm Sgn}(eV-\mu_N + U_0 + \epsilon_0 k^{n_1})
\nonumber\\
&& \times \sqrt{(eV-\mu_N+ U_0)^2- \epsilon_0^2 k^{2 n_1}}.
\label{kzeq}
\end{eqnarray}
We note that $\theta'_2$ is related to $\theta'_1$ by the relation
$2\theta'_2 = \arcsin[ \sin(2 \theta'_1)/|1-U_0/(eV-\mu_N)|]$. The
wavefunction in region II is thus given by
\begin{eqnarray}
\psi_{II}^{\rm nbs} &=& p_1 \psi_{2eR} + p_2 \psi_{2LR} + p_3
\psi_{2hL} + p_4 \psi_{2hR} \label{wavnbs2}
\end{eqnarray}

In region III, the wavefunctions constitutes a superposition of
electron-like and hole-like quasiparticles are given by
\begin{eqnarray}
\psi_{III}^{\rm nbs} = t_1 \psi^s_{3e} + t_2 \psi^s_{3h}
\label{wavnbs3}
\end{eqnarray}
where $\psi^s_{3 e(h)}$ are wavefunctions of electron- and hole-like
quasiparticles given by Eq.\ \ref{squasi1}. We note that one can
express $\theta_3^s$ and $\theta_3^{'s}$ in terms of $\theta_1$ and
$ \theta'_1$ as
\begin{eqnarray}
\sin(2 \theta_3^s) &=& \frac{\epsilon'_0 \left[\sin(2 \theta_1)
|eV+\mu_N|\right]^{n_2/n_1}}{ \epsilon_0^{n_2/n_1}\eta (\mu_S +i \zeta)} \nonumber\\
\sin(2 \theta_3^{'s}) &=& -\frac{\epsilon'_0 \sin(2 \theta'_1)
|eV-\mu_N| }{\epsilon_0^{n_2/n_1}\eta(\mu_S -i \zeta)}
\label{thetasup1}
\end{eqnarray}
Also, we find that $\psi_3^{\rm nbs} \sim e^{-i \tau_z n_1
\phi_k/2}$.

To compute $r$ and $r_A$, we need to match the boundary conditions
on the wavefunctions at $x=-d$ and $x=0$ (Fig.\ref{fig5}):
$\psi_I^{\rm nbs}(z=-d)=\psi_{II}^{\rm nbs}(z=-d)$ and
$\psi_{III}^{\rm nbs}(z=0)=\psi_{II}^{\rm nbs}(z=0)$. The boundary
condition at $z=-d$ leads to
\begin{widetext}
\begin{eqnarray}
\cos(\theta_1)e^{-i k_{z1}d} + r \sin(\theta_1) e^{i k_{z1}d} &=&
p_1 \cos(\theta_2) e^{-ik_{z2}d} + p_2 \sin(\theta_2)
e^{i k_{z2}d} \nonumber\\
\sin(\theta_1)e^{-i k_{z1}d} + r \cos(\theta_1) e^{ik_{z1}d} &=& p_1
\sin(\theta_2) e^{-i k_{z2}d} + p_2
\cos(\theta_2) e^{ik_{z2}d} \nonumber\\
r_A \cos(\theta'_1)e^{i k'_{z1}d} &=& p_4 \cos(\theta'_2)
e^{ik'_{z2}d} - p_3 \sin(\theta'_2)
e^{-i k'_{z2}d } \nonumber\\
r_A \sin(\theta'_1) e^{i k'_{z1}d } &=&  p_4 \sin(\theta'_2)
e^{ik'_{z2}d} - p_3 \cos(\theta'_2)
e^{-ik'_{z2}d} \nonumber\\
\label{supbc1}
\end{eqnarray}
while that at $z=0$ yields
\begin{eqnarray}
\left[t_1 \cos(\theta^s_3)e^{i \mu_0} + t_2
\cos(\theta_3^{'s})\right] e^{i (n_1-n_2) \phi_k/2}/\sqrt{2} &=& p_1
\cos(\theta_2) + p_2 \sin(\theta_2) \nonumber\\
\left[t_1 \sin(\theta_3^s)e^{i \mu_0} + t_2
\sin(\theta_3^{'s})\right]e^{-i (n_1-n_2) \phi_k/2}/\sqrt{2}  &=&
p_1 \sin(\theta_2)  + p_2
\cos(\theta_2)  \nonumber\\
\left[t_1 \cos(\theta_3^{s}) + t_2 \cos(\theta_3^{'s}) e^{i\mu_0}
\right]e^{i (n_1-n_2) \phi_k/2}/\sqrt{2} &=& p_4 \cos(\theta'_2) -
p_3 \sin(\theta'_2)
\nonumber\\
\left[t_1 \sin(\theta_3^s) + t_2 \sin(\theta_3^{'s}) e^{i
\mu_0}\right]e^{-i (n_1-n_2) \phi_k/2}/\sqrt{2} &=& -p_4
\sin(\theta'_2) + p_3 \cos(\theta'_2) \nonumber\\
\label{supbc2}
\end{eqnarray}
\end{widetext}

To compute the conductance, we solve for $r$ and $r_A$ numerically
using Eqs.\ \ref{supbc1} and \ref{supbc2}. One can then obtain $T_s=
(1-|r|^2+|r_A|^2)$ and obtain the tunneling conductance of the
junction using \cite{weyl1}
\begin{eqnarray}
G_s &=& G_{0s} \int_0^{k_{\rm max}^s} \frac {k dk}{2\pi}
\int_0^{2\pi} \frac{d\phi_k}{2\pi} \, T_s \label{condsup1}
\end{eqnarray}
where $k_{\rm max}^s = {\rm
Min}[(|eV+\mu_N|/\epsilon_0)^{1/n_1},(|eV-\mu_N|/\epsilon_0)^{1/n_1}]$
and $G_{0s}= (eV+\mu_N)^{2/n_1} e^2/(4 \pi h)$ is the normal state
conductance of region I. Note that the expression of $k_{\rm max}^s$
follows from the requirement that both $\sin(\theta_1)$ and
$\sin(\theta'_1)$ be real. We shall use Eq.\ \ref{condsup1} along
with Eq.\ \ref{thetasup1} for all numerical computations presented
in this section.

To make further analytical process, we now resort to the
thin-barrier limit, for which $U_0 \to \infty$ and $d \to 0$ with
$\chi= U_0 d/\hbar v_F$ held fixed. As in Sec.\ \ref{nbn}, in this
limit $\theta_2, \theta'_2, k_{z1 d}, k'_{z1}d  \to 0$ and $k_{z2}d,
[k'_{z2} d] \to -[+] \chi$. Consequently, it is easy to eliminate
$p_1$, $p_2$, $p_3$ and $p_4$ from Eqs.\ \ref{supbc1} and
\ref{supbc2}. The boundary condition in the thin barrier limit can
again be written as $\psi_I^{\rm nbs}(z=0^-) = \exp[i \tau_3 \chi]
\psi_{III}^{\rm nbs}(z=0^+)$. Thus we once again expect barrier
independence of $G$ following the same logic charted out in Sec.\
\ref{nbn}.

To verify this expectation, we first write out the above-mentioned
boundary condition equations explicitly. This leads to a set of four
equations for $r$, $r_A$, $t_1$ and $t_2$ given by
\begin{eqnarray}
&& t_1 \cos(\theta^s_3)e^{i \mu_0} + t_2 \cos(\theta_3^{'s}) =
\sqrt{2} e^{- i \alpha_0} (\cos(\theta_1) + r \sin(\theta_1))   \nonumber\\
&&t_1 \sin(\theta_3^s)e^{i \mu_0 } + t_2
\sin(\theta_3^{'s}) = \sqrt{2} e^{i \alpha_0} (\sin(\theta_1) + r \cos(\theta_1)) \nonumber\\
&&t_1 \cos(\theta_3^{s}) + t_2 \cos(\theta_3^{'s}) e^{i\mu_0} =
\sqrt{2} r_A
\cos(\theta'_1) e^{-i \alpha_0} \label{tbsupeq1}\\
&& t_1 \sin(\theta_3^s) + t_2 \sin(\theta_3^{'s}) e^{i \mu_0} =
\sqrt{2} r_A \sin(\theta'_1) e^{i \alpha_0} \nonumber
\end{eqnarray}
where $\alpha_0= (n_1-n_2)\phi_k/2 + \chi$. Solving for $r$ and
$r_A$ one obtains, in the thin barrier limit, we obtain $R^{\rm tb}=
|r|^2 =|{\mathcal Y}/{\mathcal Z}|^2$ and $R_A^{\rm tb}= |r_A|^2 =
\cos^2(2 \theta_1) |\sin(\theta_3^s-\theta_3^{'s})/{\mathcal Z}|^2$,
where ${\mathcal Y}$ and ${\mathcal Z}$ are given by
\begin{widetext}
\begin{eqnarray}
{\mathcal Y} &=& e^{2 i\mu_0} \left[ \left(e^{2 i \alpha_0}
\cos(\theta_3^s) \sin(\theta_1)- \cos(\theta_1)\sin(\theta_3^s)
\right) \sin(\theta'_1 -\theta^{'s}_3) + \left(e^{2 i \alpha_0}
\cos(\theta^{'s}_3) \sin(\theta_1)-
\cos(\theta_1)\sin(\theta^{'s}_3)
\right) \sin(\theta'_1 -\theta^s_3)  \right]\nonumber\\
{\mathcal Z} &=& \sin(\theta_1) \left[ \sin(\theta^s_3) \left(e^{2
i\mu_0} \sin(\theta'_1-\theta^{'s}_3) +
\cos(\theta'_1)\sin(\theta^{'s}_3)\right) - \sin(\theta'_1)
\cos(\theta^s_3) \sin(\theta^{'s}_3) \right]  \nonumber\\
&& - e^{2 i \alpha_0} \cos(\theta_1) \left[ \cos(\theta^s_3)
\left(e^{2 i\mu_0} \sin(\theta'_1-\theta^{'s}_3) +
\sin(\theta'_1)\cos(\theta^{'s}_3)\right) + \sin(\theta^s_3)
\cos(\theta'_1)\cos(\theta^{'s}_3) \right] \label{yzeq1}
\end{eqnarray}
\end{widetext}
We note that both $R$ and $R_A$ displays a non-trivial $\phi_k$
dependence if $n_2 \ne n_1$. Further, in the thin barrier limit, the
dimensionless barrier strength $\chi$ always appear as a phase shift
to $(n_1-n_2)\phi_k$. One can now aim to compute the transmission
$T_s^{\rm tb}$ and perform the $\phi_k$ integral. To this end, we
find, after a cumbersome calculation,
\begin{widetext}
\begin{eqnarray}
T_s^{\rm tb} &=&  (1-R^{\rm tb}+R_A^{\rm tb}) = \frac{N_s^{\rm
tb}}{D_s^{\rm tb}} = \frac{N_1 + N_2 \cos[(n_1-n_2)\phi_k +2\chi
+\beta_0] + N_3 \cos[2(n_1-n_2)\phi_k +4\chi]}{1+ D_1
\cos[(n_1-n_2)\phi_k +2\chi +\beta'_0] + D_2 \cos[2(n_1-n_2)\phi_k
+4\chi]} \label{suptranstb}
\end{eqnarray}
\end{widetext}
where $N_{1,2,3}$, $\beta_0$, $\beta'_0$, and $D_{1,2}$ are
complicated functions of $\theta_3^s$, $\theta_3^{'s}$, $\theta_1$
and $\theta'_1$. They are independent of $\chi$ and $\phi_k$;
consequently their precise forms will not be relevant for  the
subsequent discussion. In fact, from Eq.\ \ref{suptranstb}, it is
easy to check that $\int_0^{2 \pi} d\phi_k T_s^{\rm tb}$ is
independent of $\chi$ irrespective of the functional forms of
$N_{1,2,3}$, $\beta_0$, $\beta'_0$ and $D_{1,2}$. The simplest way
to see this is to use the standard substitution $z=
\exp[i\{(n_1-n_2)\phi_k +2 \chi \}]$ and convert the integral over
$\phi_k$ to a complex integral over unit circle. The denominator,
written in terms of $z$, is a quartic polynomial in $z$ leading to
four poles inside the unit circle. The residues of these poles do
not depend on $\chi$. Thus we expect that $G_s$ will be independent
of $\chi$ in the thin barrier limit. We once again note that as in
NBN junctions, the $\chi$ independence is a consequence of change in
the topological winding number across the junction; $G_s$ will be an
oscillatory function of $\chi$ if $n_1=n_2$.

\begin{figure}
{\includegraphics[width=0.48\hsize]{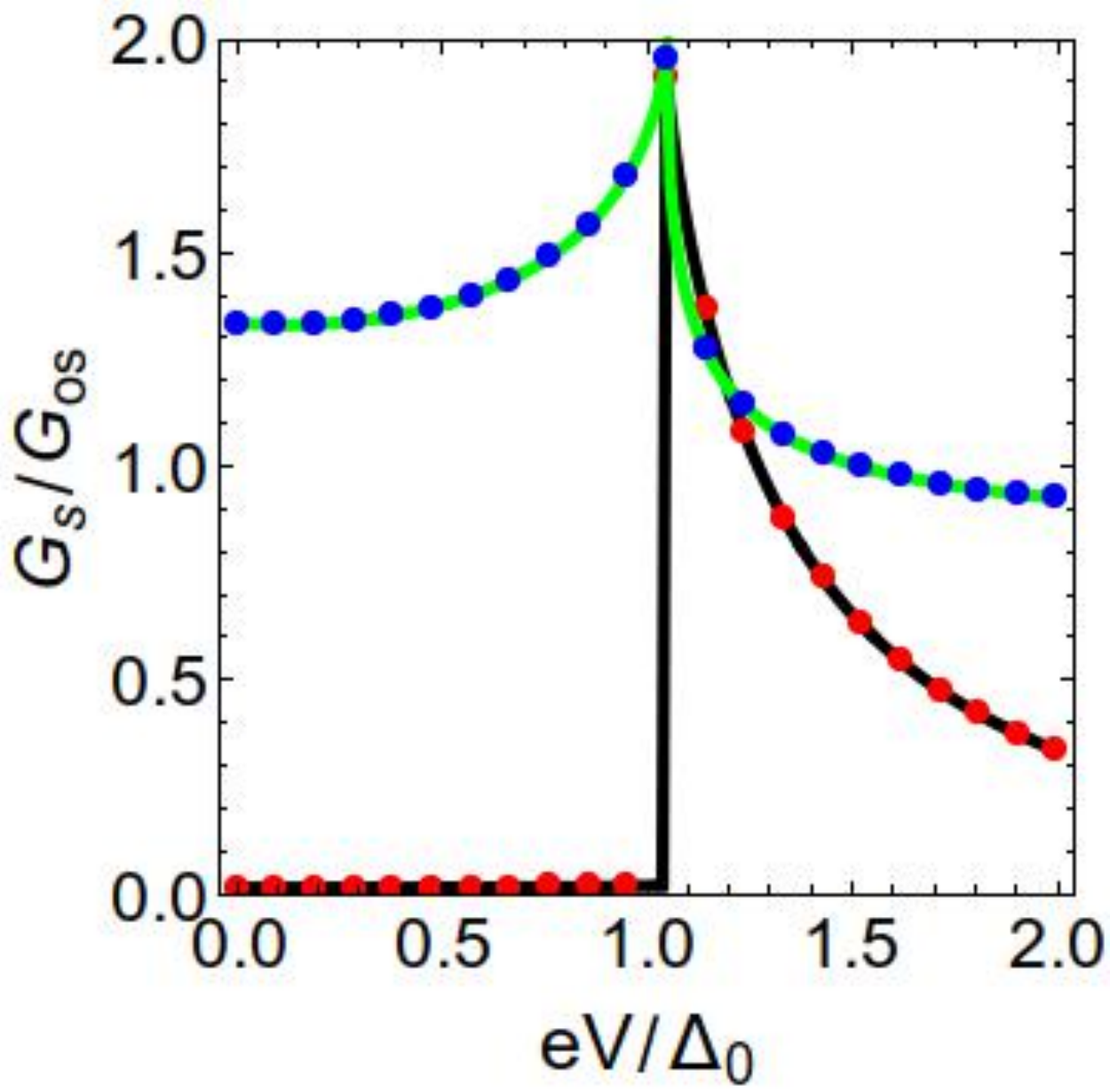}}
{\includegraphics[width=0.48 \hsize]{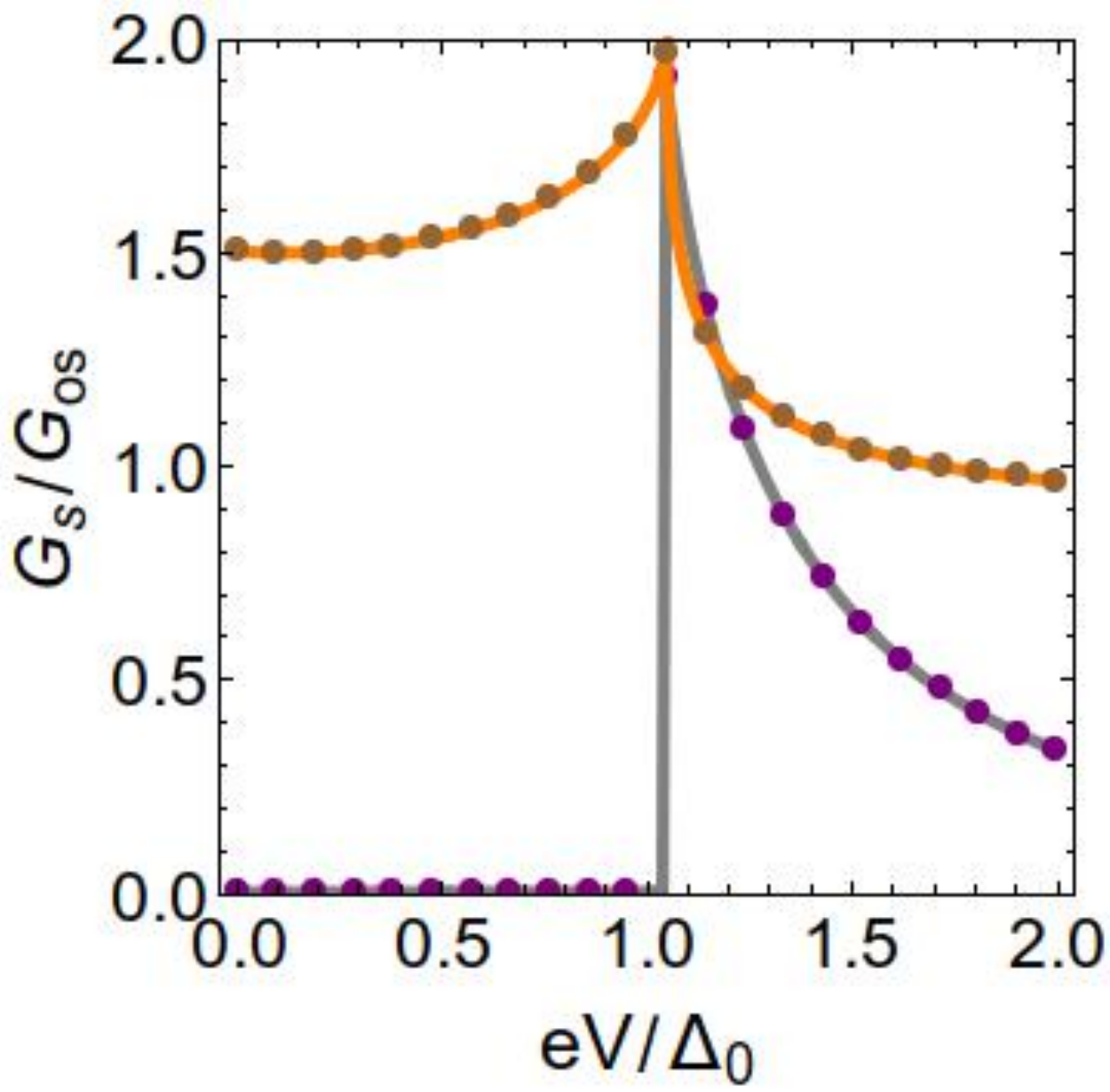}} \caption{ (a) Plot of
$G_s/G_{0s}$ as a function of $eV/\Delta_0$ for $\delta n=\pm 1$.
The black solid line (red dots) corresponds to $n_2=2$, $n_1=1$, and
$\chi=0 (\pi/4)$ while the blue solid line (green dots) corresponds
to $n_2=1$, $n_1=2$ and $\chi=0(\pi/4)$. (b) Plot of $G_s/G_{0s}$ as
a function of $eV/\Delta_0$ for $\delta n=\pm 2$. The grey solid
line (pink dots) corresponds to $n_2=3$ and $n_1=1$ and $\chi=0
(\pi/4)$ while the orange solid line (brown dots) corresponds to
$n_2=1$, $n_1=3$ and $\chi=0(\pi/4)$ For all plots,
$\epsilon_0=\epsilon'_0=\eta=1$ and $\mu_N=\mu_S=100 \Delta_0$. See
text for details. \label{fig6}}
\end{figure}

\begin{figure}
{\includegraphics[width=0.48\hsize]{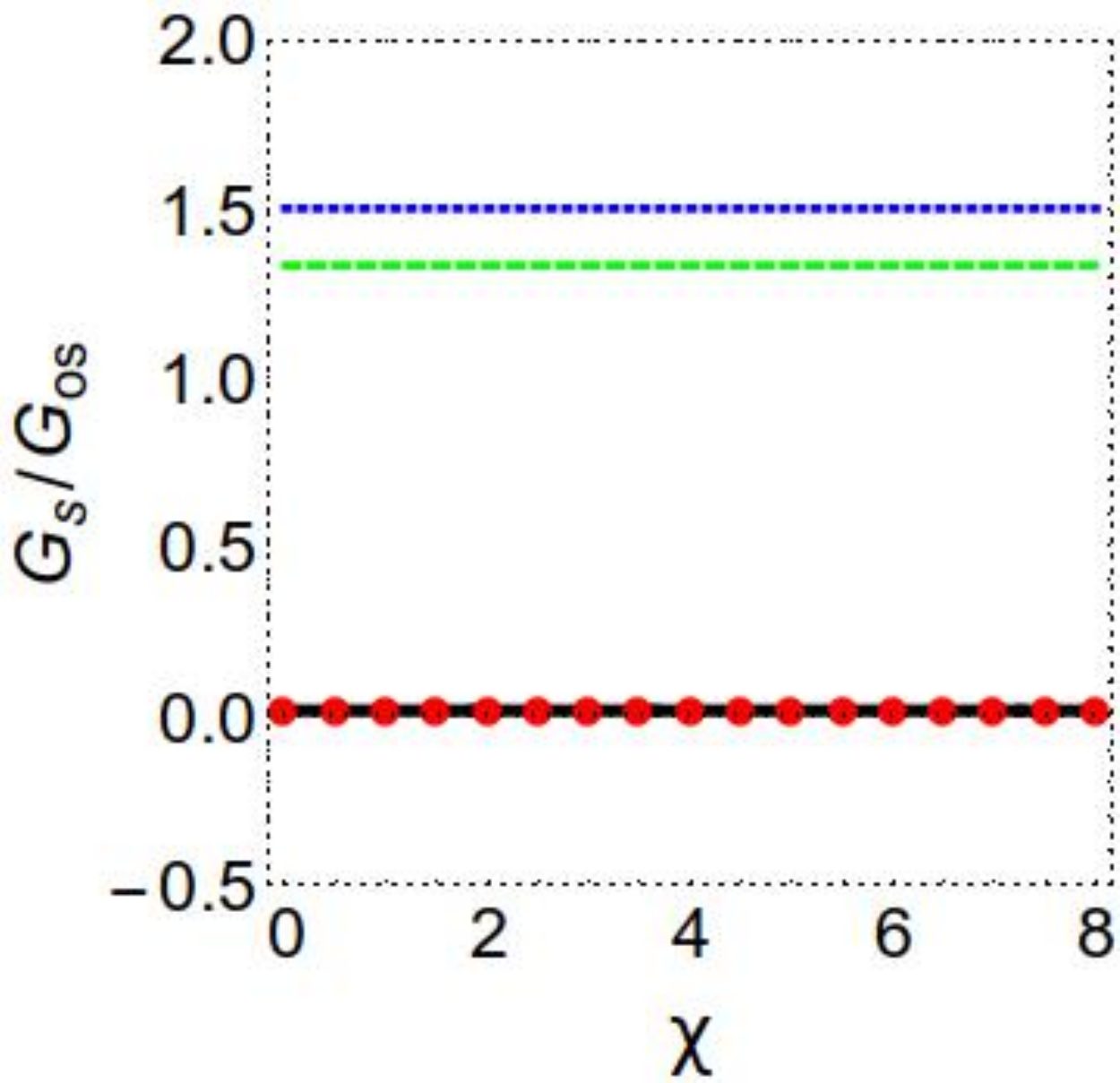}}
{\includegraphics[width=0.48 \hsize]{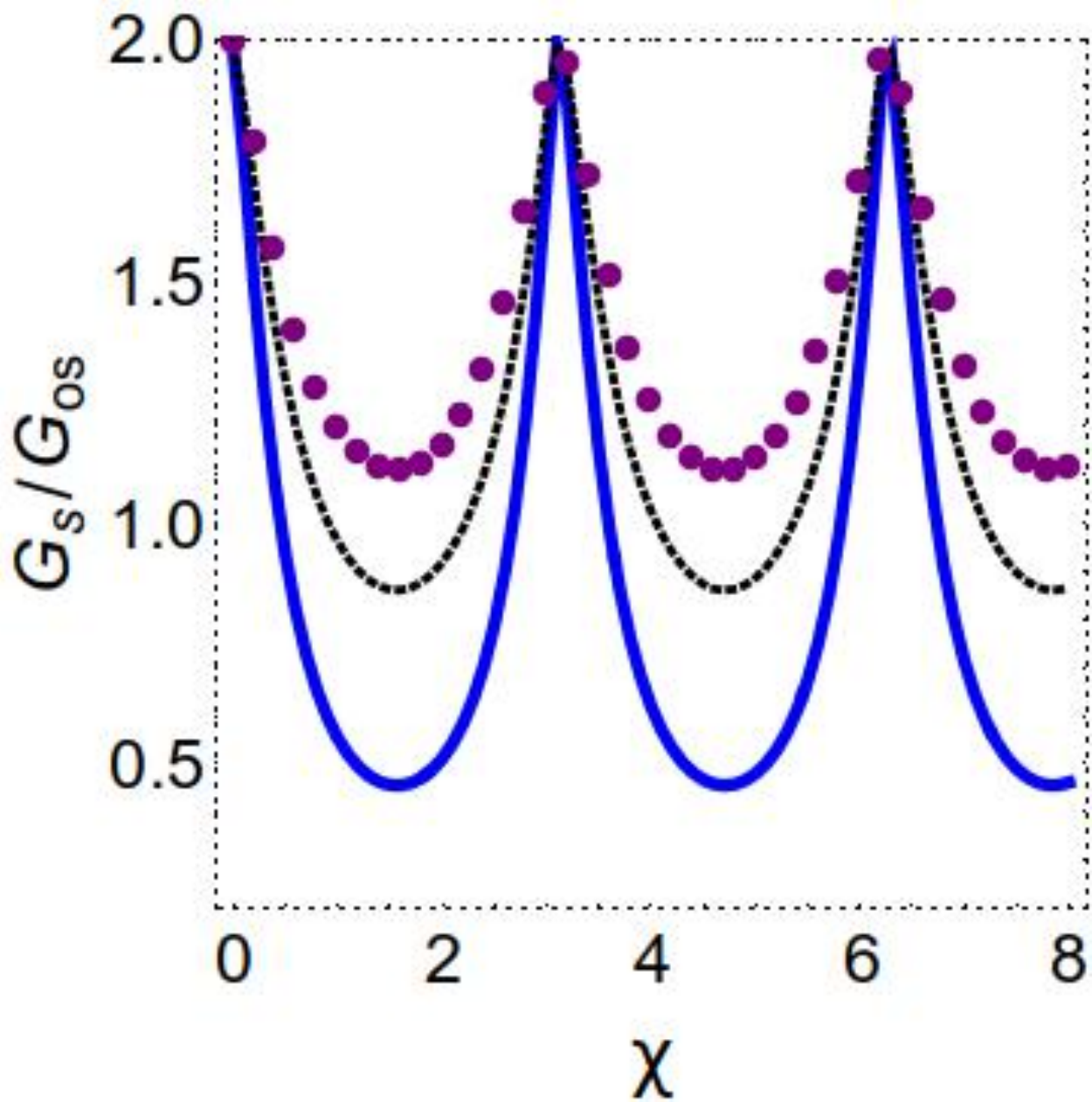}} \caption{ (a) Plot of
zero-bias conductance $G_s(eV=0)/G_{0s}$ as a function of $\chi$.
The black solid line (red dots) correspond to $n_1=1$ and $n_2=2(3)$
while the blue solid (green dashed) lines correspond to $n_2=1$ and
$n_1=2(3)$. (b) Plot of $G(eV=0)/G_{0s}$ as a function of $\chi$ for
$n_1=n_2=n_0$. The blue solid, the black dashed, and the magenta
dotted lines correspond to $n_0=1$, $2$ and $3$ respectively. All
other parameters are same as in Fig.\ \ref{fig6}. See text for
details. \label{fig7}}
\end{figure}

The qualitative reasoning presented above can be supported by
numerics in the thin barrier limit presented Figs.\ \ref{fig6} and
\ref{fig7}. From Fig.\ \ref{fig6}, we find that the subgap tunneling
conductance vanishes for $n_2 >n_1$ but remains finite for $n_2 \le
n_1$. Moreover $G_s$ is independent of $\chi$. This barrier
independence is further highlighted in Fig.\ \ref{fig7}, where we
find that the zero-bias conductance ($G_s(eV=0)$) becomes
independent of $\chi$ in the thin barrier limit for $n_1 \ne n_2$;
in contrast for $n_1=n_2$, a clear oscillatory behavior is found.

The suppression of $G_s$ for $eV \le \Delta_0$ and $n_2 > n_1$ in
Figs.\ \ref{fig6}(a) and \ref{fig6}(b) can be qualitatively
understood from Eq.\ \ref{thetasup1}. We first note that our
numerical results for the thin barrier limit are presented for
$\mu_s =\mu_N \gg eV, \Delta_0$ and $\epsilon_0=\epsilon'_0=\eta=1$.
In this limit, one finds, from Eq.\ \ref{thetasup1},
$\sin(2\theta_3^s) \simeq (\sin(2\theta_1))^{n_2/n_1}
\mu_N^{n_2/n_1-1}$. Thus for $eV \le \Delta_0$, $\theta_3^s$ has no
real solution for a majority of the transverse channels for which
$(\sin(2\theta_1))^{n_2/n_1} > \mu_N^{1-n_2/n_1}$. For these
transverse modes, $\sin(2\theta_3^s), \cos(2\theta_3^s) \to i
\exp[(\sin(2\theta_1))^{n_2/n_1}\mu_N^{n_2/n_1-1}]/2$ for large
$\mu_N$. A similar behavior is found for $\theta_3^{'s}$. Thus from
Eq.\ \ref{yzeq1} one finds that for these modes ${\mathcal Z} \sim
\exp[ (\sin(\theta_1))^{n_2/n_1} \mu_N^{n_2/n_1-1}]$. Thus $R_A \sim
\sin^2(\theta_3^s-\theta_3^{'s})/|{\mathcal Z}|^2 \sim \exp[-
\mu_N^{n_2/n_1-1} (\sin(2\theta_1))^{n_2/n_1}]$ and vanishes
exponentially for these modes. It is also easy to see that $R \to 1$
for these modes. The number of such modes constitute a majority of
the total available transverse modes for large $\mu_N$;
consequently, $G_s(eV\le \Delta_0) \to 0$ in this limit. We note
that the suppression of the subgap tunneling conductance for large
$\mu_N$ and $\mu_S$ is completely controlled by the change of the
topological winding numbers $n_1$ and $n_2$ across the junctions. In
contrast, for $n_2 < n_1$, $\theta_3^s \to 0$, since $\mu_N^{n_2/n_1
-1} \ll 1$ in this regime. Similarly, from Eq.\ \ref{thetasup1}, we
find that $\theta_3^{'s} \to \pi/2$ in this limit. Thus $R_A$ remain
finite and one finds finite subgap $G_s$ as can be seen in Fig.\
\ref{fig6}. Thus we conclude that the subgap tunneling conductance
of these junctions depends crucially on the ratio $n_2/n_1$. We note
that for $n_1=1,n_2=1$, our results reproduces those in Ref.\
\onlinecite{weyl3} for $\mu_N, \mu_s \gg \Delta_0$ as special case.

\begin{figure}
{\includegraphics[width=0.98\hsize]{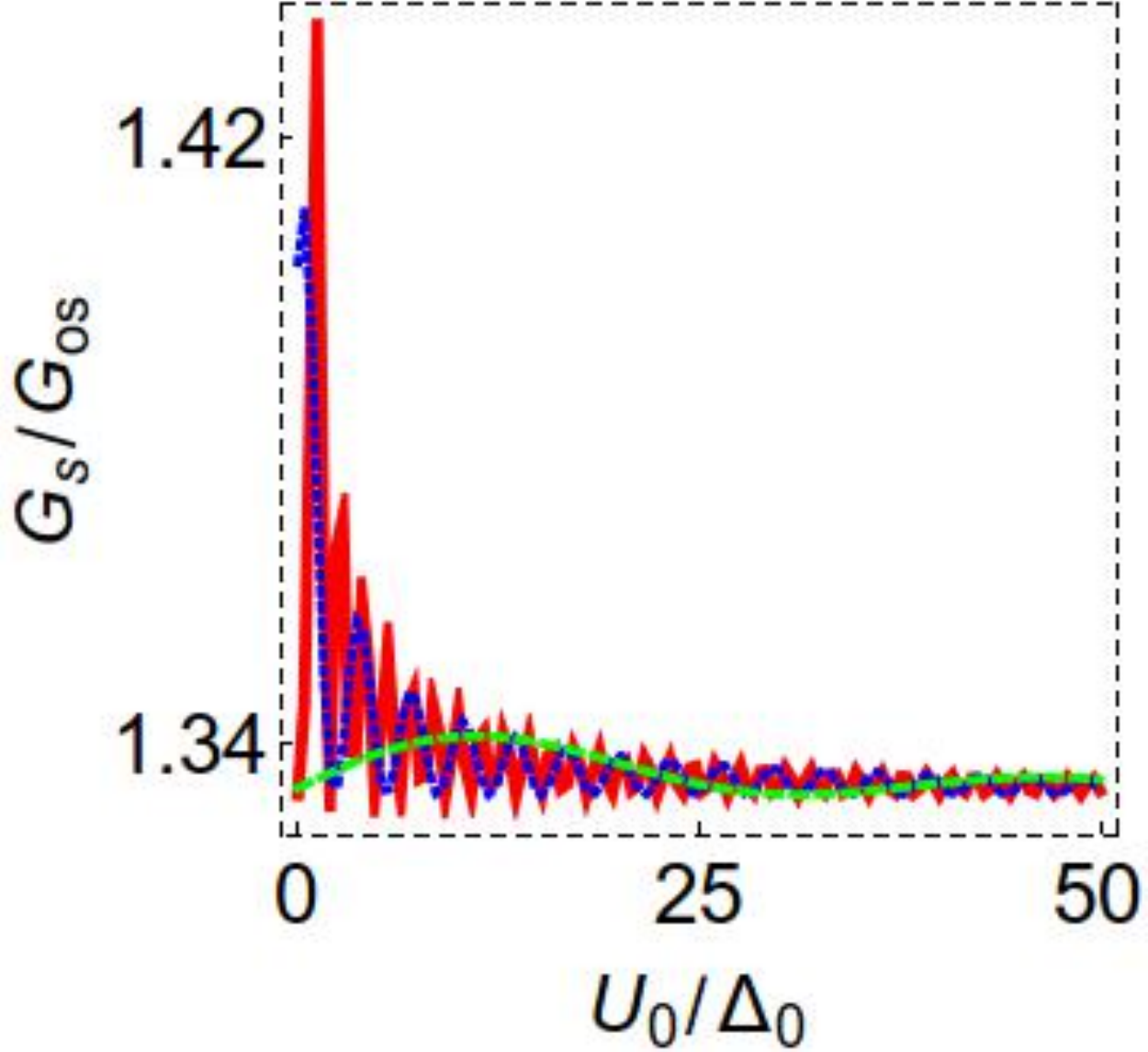}} \caption{ (a) Plot of
zero-bias conductance $G_s(eV=0)/G_{0s}$ as a function of
$U_0/\Delta_0$ for $n_1=2$, $n_2=1$, and $\mu_N=0.1 \Delta_0$ and
$\mu_S=100 \Delta_0$. The red solid line correspond to $dk_0=10$,
the blue dotted line to $dk_0=1$, and the green dashed line to $d
k_0=0.1$. See text for details. \label{fig8}}
\end{figure}

Finally, we consider deviation from the thin barrier limit. To this
end, we numerically evaluate the conductance using Eqs.\
\ref{supbc1}, \ref{supbc2}, and \ref{condsup1} and plot the
zero-bias conductance $G_s(eV=0)/G_{0s}$ as a function of
$U_0/\Delta_0$ for several representative values of $d$ in Fig.\
\ref{fig8}. As in the case of NBN junctions, we find that
$G_{s}(eV=0)/G_{0s}$ shows oscillatory behavior for large $d$;
however the amplitude of these oscillations decay with increasing
$U_0$ and approaches the thin barrier behavior for either large
$U_0/\Delta_0$ or small $d$. We note that even for large $d$, a
sufficiently large value of $U_0/\Delta_0$ can lead to near constant
behavior of $G_s$ which is expected to make this behavior
experimentally easy to verify.

\section{Discussion}
\label{diss}

In this work, we have studied the tunneling conductance between
junctions of a WSM and a MSM ( or two MSMs) where the topological
winding numbers of the Weyl nodes change across the junction. We
have shown that the tunneling conductance of such junctions exhibits
several unconventional features which are absent both in junctions
involving 2D topological materials such graphene or topological
insulators surfaces and in those made out of 3D topological
materials such as WSMs or MSMs with $n_1=n_2$. The most striking of
such features is the barrier independence of $G$ and $G_s$ in the
thin barrier limit. We note that such a feature is in sharp contrast
to both Schrodinger materials (where $G$ decays monotonically with
increasing $\chi$) and previously studied topological materials
(where $G$ oscillates with $\chi$). We demonstrate that such barrier
independence is a consequence of the change in topological winding
number of the Weyl nodes across the junction. Moreover, for NBS
junctions with $\mu_S,\mu_N \gg eV,\Delta_0$, the subgap tunneling
conductance $G_s(eV\le \Delta_0)$ vanishes when $n_2>n_1$; however,
it is finite when $n_1 >n_2$. Thus the subgap tunneling conductance
of such NBS junctions depend crucially on the ratio of the
topological winding numbers of the WSMs/MSMs forming the junction.

The simplest experimental verification of our work would require
formation of a junction between a WSM and MSM. The longitudinal
direction of such junctions needs to be the symmetry axis of the MSM
(taken to be $\hat z$ in our work). The barrier regions can be
simulated by putting an additional local gate voltage $U_0$ on the
WSM in a region of width $d$. For large $U_0$, we predict that
$G(eV)$ will be independent of the dimensionless barrier strength
$\chi$. Another, experimentally more challenging, possibility would
be to study the subgap tunneling conductance of such junctions when
superconductivity is induced either on the WSM or the MSM. We
predict that the subgap tunneling conductance $G_s(eV \le \Delta_0)$
in these two cases will show qualitatively different behavior for
$\mu_S,\mu_N \gg eV, \Delta_0$. For the case, when superconductivity
is induced in the MSM, $G_s$ will vanish; in contrast it will be
finite, if superconductivity is induced in the WSM. However, in both
cases, $G_s$ will be independent of $\chi$ for large $U_0$. We note
that such features can also be observed in a junction constructed
out of two MSMs of similar material provided one applies a
sufficiently large strain on one of them \cite{msm2}. This would
split the Weyl nodes leading to $n=1$ in that region while the other
region of the junction will still have $n \ne 1$. This will lead to
the crucial jump in topological winding number across the junction
and lead to predicted the barrier independent transport.

In conclusion, we have studied transport in NBN and NBS junctions
between a WSM and a MSM or two MSMs with different topological
winding numbers. We have demonstrated barrier independence of
tunneling conductance for such junctions in the thin barrier limit
and analyzed the role of the topological winding numbers in shaping
the applied voltage dependence of their tunneling conductance. We
have discussed experimental signatures of these phenomena.

{\it Acknowledgement:} KS thanks Pushan Majumdar, Koushik Ray, and
Diptiman Sen for several discussions.

\appendix

\section{Inter-node scattering}

In this appendix, we discuss the effect of inter-node scattering on
the conductivity calculations for NBN junctions. First, we note that
since the potential barrier conserves transverse momentum on
scattering such inter-node scattering can only occur if the Weyl
nodes occur at same transverse momentum. To this end, we consider a
model Hamiltonian with two Weyl nodes at $(0,0,\pm K_0)$ which is
given by
\begin{eqnarray}
H_w &=& E_0 \sum_{\vec k} \psi_{\vec k}^{\dagger} ( (k_z^2- K_0^2)
\tau_z + \nonumber\\
&& \alpha_{n_1} k^{n_1} \left[\cos(n_1 \phi_k) \tau_x + \tau_y
\sin(n_1 \phi_k) \right] ) \psi_{\vec k} \label{hw1}
\end{eqnarray}
where $E_0 = \hbar^2/(2m a^2)$ is the unit of energy, $a$ is the
lattice spacing, all $k_{x,y,z}$ are measured in units of $a^{-1}$,
$\phi_k= \arctan[k_y/k_x]$,  $\alpha_{n_1}$ is a material specific
constant, $k=\sqrt{k_x^2+k_y^2}$ is the magnitude of the transverse
momentum, and $n_1$ is the topological winding number of the nodes.
In what follows we are going to study transport through an NBN
junction whose basic quasiparticle excitations are governed by $H_w$
allowing for inter-node scattering between the two Weyl nodes
\cite{weyl4}. The schematic picture of such a junction is shown in
Fig.\ \ref{fig1} of the main text. The analysis carried out here
will be similar to that in the main text and we are going to present
the salient features which are different due to the presence of
inter-node scattering. We note that the energy dispersion of $H_w$
is given by
\begin{eqnarray}
E_w= \pm \sqrt{(k_z^2-K_0^2)^2 + \alpha^2_{n_1} |k|^{2n_1}}
\label{entn1}
\end{eqnarray}

The wavefunction in region I is now a superposition of that of an
incident electron at the node $(0,0, K_0)$ ( which we shall denote
as node 1) with momentum $k_z^{+}$ and two reflected electrons at
the two nodes with momenta $k_z^{-}$ (intra-node) and $-k_z^{+}$
(inter-node). The expressions for these momenta and the
corresponding electron wavefunctions can be easily found using Eqs.\
\ref{hw1} and \ref{entn1} and is given by
\begin{eqnarray}
k_z^{\pm} &=& \sqrt{ K_0^2 \pm \sqrt{(eV+\mu_N)^2-\alpha_{n_1}^2
k^{2n_1}}} \nonumber\\
\psi_{\rm in 1}  &=& e^{ i (k_z^{+} z + k_x x + k_y y -n_1 \sigma_z
\phi_{\vec k}/2)} \frac{1}{\sqrt{1+\eta_1^2}} \left (
\begin{array}{c} 1 \\ \eta_1 \end{array} \right) \nonumber\\
\psi_{\rm ref1} &=& e^{ i (k_z^{-} z + k_x x + k_y y -n_1 \sigma_z
\phi_{\vec k}/2)} \frac{1}{\sqrt{1+\eta_1^2}} \left (
\begin{array}{c} \eta_1 \\ 1 \end{array} \right) \nonumber\\
\psi'_{\rm ref2} &=& e^{ i (-k_z^{+} z + k_x x + k_y y -n_1 \sigma_z
\phi_{\vec k}/2)} \frac{1}{\sqrt{1+\eta_1^2}} \left (
\begin{array}{c} 1 \\ \eta_1 \end{array} \right)  \label{eifntn1}
\end{eqnarray}
Here $\psi_{\rm ref1}$ ($\psi_{\rm ref2}$)are the wavefunctions of
the electrons reflected at the same (opposite) node, $\eta_1=
\arcsin[\alpha_{n_1} k^{n_1}/(eV+\mu_N)]$, and we have chosen the
energy of the incident electron to be $\epsilon= eV+\mu_N$. Thus the
wavefunction in region I is given by
\begin{eqnarray}
\psi_I &=& \psi_{\rm in 1} + r_1 \psi_{\rm ref1} + r_2 \psi_{\rm
ref2} \label{wavtn1}
\end{eqnarray}
where $r_1$ and $r_2$ re the amplitudes of intra- and inter-node
reflection respectively.

In region II, the wavefunction is a linear superposition of left and
right moving electrons in both nodes. The wavefunctions for the
right moving electrons in the node situated in $(0,0,K_0)$ and the
left moving electrons in both nodes can be read off from Eq.\
\ref{eifntn1}. Indeed their expressions are given by Eq.\
\ref{eifntn1} with $k_z^{\pm} \to k_z^{'\pm} = \sqrt{ K_0^2 \pm
\sqrt{(eV+\mu_N-U_0)^2-\alpha_{n_1}^2 k^{2n_1}}}$ and $\eta_1 \to
\eta_2= \arcsin[\alpha_{n_1} k^{n_1}/(eV+\mu_N-U_0)]$. We denotes
these wavefunctions by $\psi'_{\rm in 1}$, $\psi'_{\rm ref 1}$ and
$\psi'_{\rm ref 2}$. The wavefunction for the right moving electron
around the second Weyl node $(0,0,-K_0)$ ( which we shall denote as
node 2) is given by
\begin{eqnarray}
\psi'_{\rm in 2} &=& e^{ i (-k_z^{'-} z + k_x x + k_y y -n_1
\sigma_z \phi_{\vec k}/2)} \frac{1}{\sqrt{1+\eta_2^2}} \left (
\begin{array}{c} \eta_2 \\ 1 \end{array} \right) \label{eifntn2}
\end{eqnarray}
The wavefunction in region II can be written in terms of these
wavefunctions as
\begin{eqnarray}
\psi_{II} = p_1 \psi'_{\rm in 1} + p_2 \psi'_{\rm in 2} + q_1
\psi'_{\rm ref 1} + q_2 \psi'_{\rm ref 2} \label{wavtn2}
\end{eqnarray}
where $p_{1,2}(q_{1,2})$ are the amplitudes of right (left) moving
electrons in nodes $1$ or $2$.

In region III, the wavefunction is a linear combination of
right-moving electron wavefunctions on both nodes. These are obtain
from Eq.\ \ref{eifntn1} and \ref{eifntn2} with the substitution of
$\mu_N \to \mu'_N$, $n_1 \to n_2$, $k_z^{\pm} \to k_{3z}^{\pm} =
\sqrt{ K_0^2 \pm \sqrt{(eV+\mu'_N)^2-\alpha_{n_2}^2 k^{2n_2}}}$ and
$\eta_1 \to \eta_3= \arcsin[\alpha_{n_2}k^{n_2}/(eV+ \mu'_N)]$. We
denotes these wavefunctions by $\psi_{3}$ and $\psi'_{3}$. The
wavefunction in this region are given by
\begin{eqnarray}
\psi_{III} = t \psi_{3} + t' \psi'_3 \label{wavtn3}
\end{eqnarray}
where $t$  and $t'$ denote amplitudes of transmission of electrons
in node $1$ and node $2$ respectively.

The current conservation at the boundaries, {\it i.e.}, at $z=0$ and
$z=d$, requires the continuity of the wavefunction and their $z$
derivatives at $z=0$ and $z=d$. The conditions $\psi_I(z=0)=
\psi_{II}(z=0)$ and $\partial_z \psi_I(z=0) = \partial_z
\psi_{II}(z=0)$ yields
\begin{widetext}
\begin{eqnarray}
\frac{1}{\sqrt{1+ \eta_1^2}} ( 1+ r_1 + \eta_1 r_2) &=&
\frac{1}{\sqrt{1+ \eta_2^2}} ( p_1+ q_1 + \eta_2(p_2+ q_2))
\nonumber\\
\frac{1}{\sqrt{1+ \eta_1^2}} ( \eta_1(1+ r_1) + r_2) &=&
\frac{1}{\sqrt{1+ \eta_2^2}} ( \eta_2(p_1+ q_1) + p_2+ q_2)
\nonumber\\
\frac{1}{\sqrt{1+ \eta_1^2}} ( k_z^{+} (1- r_1) + \eta_1 k_z^- r_2)
&=& \frac{1}{\sqrt{1+ \eta_2^2}} ( (p_1- q_1)k_z^{'+} - \eta_2
k_z^{'-}(p_2- q_2)) \nonumber \\
\frac{1}{\sqrt{1+ \eta_1^2}} ( \eta_1 k_z^{+} (1- r_1) + k_z^- r_2)
&=& \frac{1}{\sqrt{1+ \eta_2^2}} ( (p_1- q_1) \eta_2 k_z^{'+} -
k_z^{'-}(p_2- q_2)) \label{z0cond}
\end{eqnarray}
\end{widetext}
Similarly the conditions  $\psi_{II}(z=d) = \psi_{III}(z=d)$ and
$\partial_z \psi_{II}(z=d) = \partial_z \psi_{III}(z=d)$ yields
\begin{widetext}
\begin{eqnarray}
\frac{1}{\sqrt{1+ \eta_3^2}} [ t e^{ik_z^+ d} +  \eta_3 t' e^{ik_z^-
d}] &=& \frac{1}{\sqrt{1+ \eta_2^2}} ( p_1 e^{i k_z^{'+} d} + q_1
e^{- i k_z^{'+} d} + \eta_2(p_2 e^{- i k_z^{'-} d} + q_2 e^{i
k_z^{'+} d} )) e^{-i \nu'}
\nonumber\\
\frac{1}{\sqrt{1+ \eta_3^2}} [ t \eta_3 e^{ik_z^+ d} +  t' e^{ik_z^-
d}] &=& \frac{1}{\sqrt{1+ \eta_2^2}} ( \eta_2(p_1 e^{i k_z^{'+} d} +
q_1 e^{- i k_z^{'+} d}) + p_2 e^{- i k_z^{'-} d} + q_2 e^{i k_z^{'+}
d}) e^{i \nu'} \nonumber\\
\frac{1}{\sqrt{1+ \eta_3^2}} [ t k_z^+ e^{ik_z^+ d} +  \eta_3 k_z^-
t' e^{ik_z^- d}] &=& \frac{1}{\sqrt{1+ \eta_2^2}} ( k_z^{'+} (p_1
e^{i k_z^{'+} d} - q_1 e^{- i k_z^{'+} d}) -k_z^{'-} \eta_2(p_2 e^{-
i k_z^{'-} d} - q_2 e^{i k_z^{'+} d} )) e^{-i \nu'}\nonumber\\
\frac{1}{\sqrt{1+ \eta_3^2}} [ t \eta_3 k_z^+ e^{ik_z^+ d} + k_z^-
t' e^{ik_z^- d}] &=& \frac{1}{\sqrt{1+ \eta_2^2}} ( \eta_2 k_z^{'+}
(p_1 e^{i k_z^{'+} d} - q_1 e^{- i k_z^{'+} d}) -k_z^{'-} (p_2 e^{-
i k_z^{'-} d} - q_2 e^{i k_z^{'+} d} )) e^{i \nu'} \label{zdcond}
\end{eqnarray}
\end{widetext}
where $\nu'= (n_2-n_1)\phi_k/2$. In what follows, we shall
numerically solve Eq.\ \ref{z0cond} and \ref{zdcond} to obtain $r_1$
and $r_2$. The conductance can then be computed as
\begin{eqnarray}
G(eV) &=&  2G_0 \int d^2 k_y T(k_x, k_y; eV), \nonumber\\
T(k_x,k_y;eV) &=& (1-|r_1|^2v_{z2}/v_{z1}-|r_2|^2)  \label{condtn}
\end{eqnarray}
where $G_0$ is defined in the main text with $n_0=2$ and $v_{z 1,2}$
are the longitudinal velocities of the Weyl fermions in the two
nodes. Comparing Eq.\ \ref{condtn} with Eq.\ \ref{condrel1} of the
main text, we find that they coincide when $|r_1|^2 \to 0$. This is
also the regime where the approximation used in the main text should
work.
\begin{figure}
{\includegraphics[width=0.98\hsize]{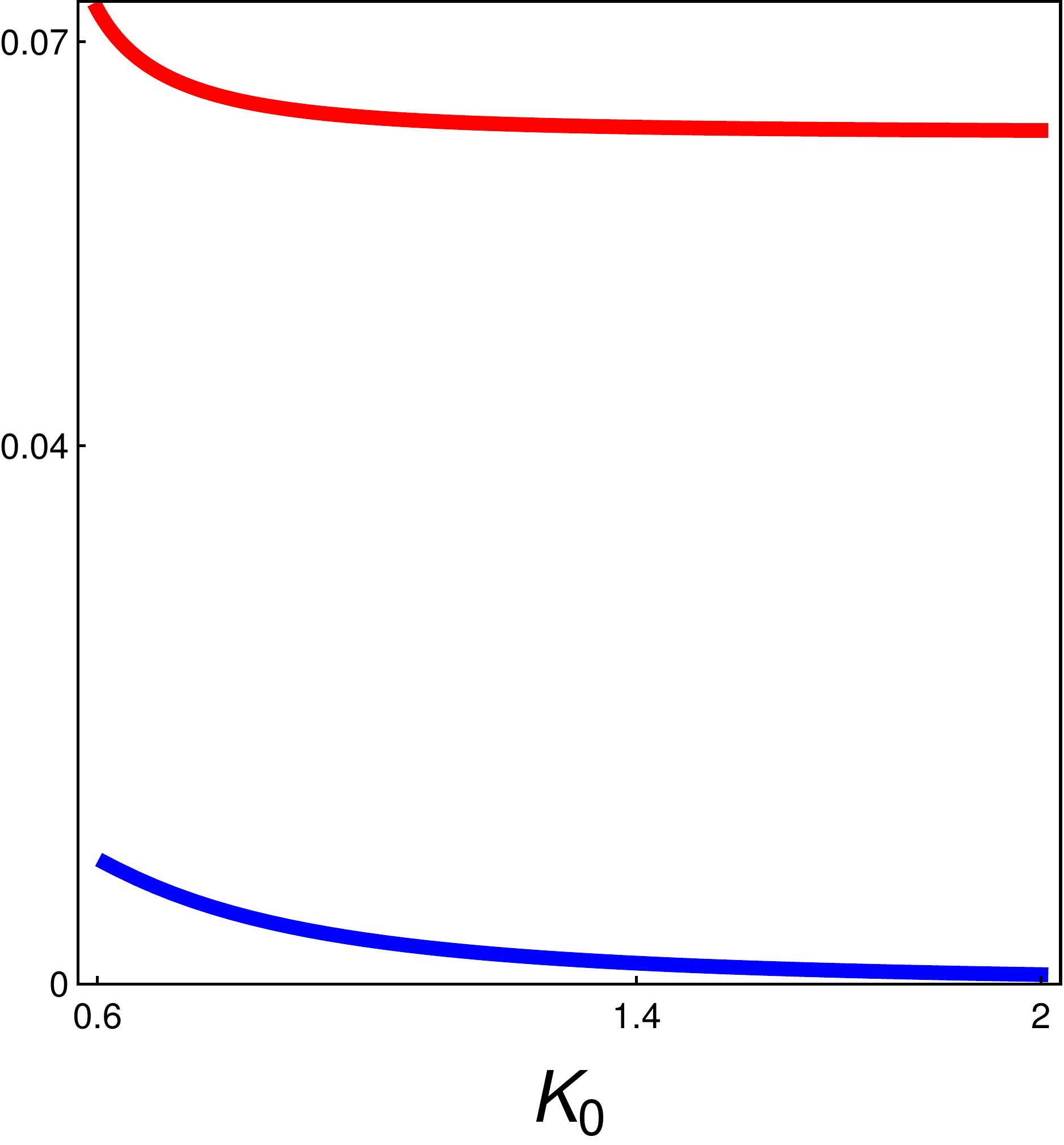}} \caption{ Plot of
inter-node scattering probability $R_{1,2} = \int d^2 k |r_{1,2}|^2$
as a function of $K_0$ (all momenta are in units of $a$). The blue
solid line represents the inter-node scattering probability $R_1$
and the red line $R_2$. Here $eV=0.01$, $\mu_L=0.1$, $\mu_R=0.3$,
$U_0=0.5$, $n_1=n_2=1$, and $d=0.2$. All energy scales are in units
of $\hbar^2/(2 ma^2)eV=0.01$. \label{figa1}}
\end{figure}

\begin{figure}
{\includegraphics[width=0.98\hsize]{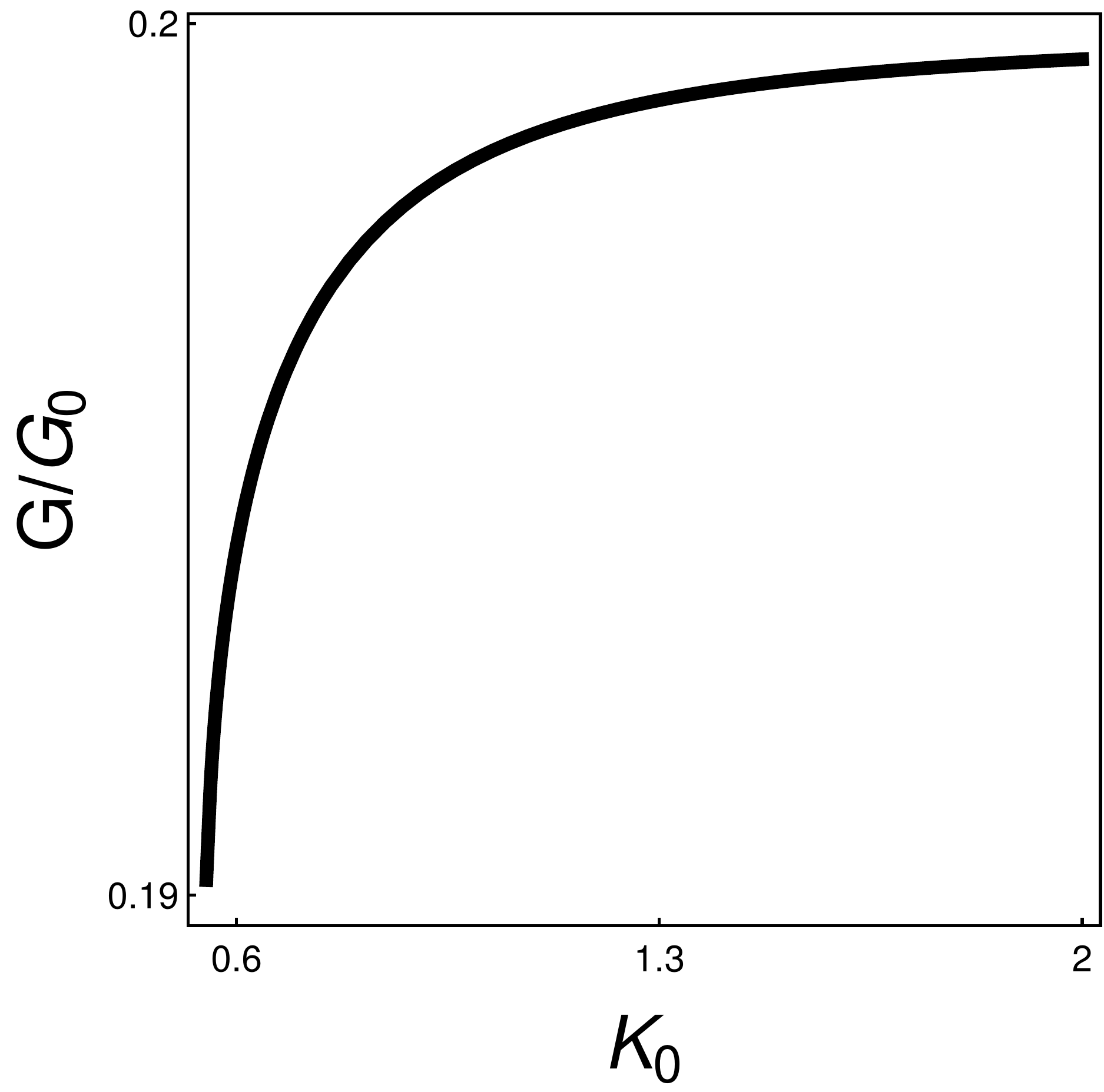}} \caption{ Plot of the
conductance $G/G_0$ as a function of $K_0$ (all momenta are in units
of $a$). All parameters are same as in Fig.\ \ref{figa1}.
\label{figa2}}
\end{figure}

The behavior of the $R_{1,2} = \int d^2 k |r_{1,2}|^2$ is shown in
Fig.\ \ref{figa1} for a representative value of $\mu_R$, $d$, $eV$,
$U_0$ and $\mu_L$. We note that for large enough $K_0$, the
inter-valley scattering is largely suppressed and one finds that for
$K_0 a \ge 0.5$, the inter-node scattering can be safely neglected
compared to the intra-node scattering. This behavior is seen for all
parameter values that we have checked. Thus we note that as long as
the Weyl nodes are far off in momentum space, for low-energy
transport, it is possible to work within independent node
approximation as done in the main text. Indeed as seen from the
behavior of $G/G_0$ in Fig.\ \ref{figa2}, $G/G_0$ becomes almost
independent of $K_0$ in this limit. Thus we expect that for large
enough $K_0 a>0.5$, inter-node scattering do not affect conductance.
Indeed, comparing $G/G_0$ in Fig.\ \ref{fig2}, to $G/G_0$ computed
by setting $T=1-|r_2|^2$ (i.e. ny setting $r_1=0$) yields a
near-perfect match.

Next, we consider the thin barrier limit in these junctions. This
limit is given by $E_0 K_0^2 a^2 \gg U_0 \gg \mu_N+ eV$. In this
limit one finds $\chi= U_0 d/(2 \hbar v_F K_0) \ll K_0 d$,
$k_{z2}^{'\pm} d = K_0 d \pm \chi$, $\eta_2 \to 0$, and $k_{3z} d ,
k'_{3z} d \to 0$. In this limit, it is easy to see that Eqs.\
\ref{z0cond} and \ref{zdcond} reduces to
\begin{widetext}
\begin{eqnarray}
\frac{1}{\sqrt{1+ \eta_1^2}} ( 1+ r_1 + \eta_1 r_2) &=& p_1+ q_1
\nonumber\\
\frac{1}{\sqrt{1+ \eta_1^2}} ( \eta_1(1+ r_1) + r_2) &=& p_2+ q_2
\nonumber\\
\frac{1}{\sqrt{1+ \eta_1^2}} ( k_z^{+} (1- r_1) + \eta_1 k_z^- r_2)
&=&  (p_1- q_1)K_0 \nonumber \\
\frac{1}{\sqrt{1+ \eta_1^2}} ( \eta_1 k_z^{+} (1- r_1) + k_z^- r_2)
&=& K_0(p_2- q_2)) \nonumber\\
\frac{1}{\sqrt{1+ \eta_3^2}} (t +  \eta_3 t' ) &=&[ p_1 e^{i K_0 d}
+ q_1 e^{- i K_0 d} ]e^{-i \nu }
\nonumber\\
\frac{1}{\sqrt{1+ \eta_3^2}} [ t \eta_3 +  t' ] &=& ( p_2 e^{- i K_0
d} + q_2 e^{i K_0 ^{'+} d}) e^{i \nu } \nonumber\\
\frac{1}{\sqrt{1+ \eta_3^2}} [ t k_z^+  +  \eta_3 k_z^- t'] &=& (
K_0 (p_1 e^{i K_0 d} - q_1 e^{- i K_0 d})) e^{-i \nu}\nonumber\\
\frac{1}{\sqrt{1+ \eta_3^2}} [ t k_z^{+} \eta_3 + k_z^- t'] &=&
\frac{-1}{\sqrt{1+ \eta_2^2}} ( K_0 (p_2 e^{- i K_0 d} - q_2 e^{i
K_0 d} )) e^{i \nu } \label{zcondtb}
\end{eqnarray}
\end{widetext}
where $\nu= (n_2-n_1)\phi_k/2 + \chi$ and we have assumed that $2
\chi << K_0 d$. Thus we find that in this limit, the barrier
potential once again appears as a constant shift in $\phi_k$ as can
be seen from expression of $\nu$; consequently, $G$ becomes
independent of $\chi$ if $n_2 \ne n_1$. Thus we expect to reproduce
the results of the main text in this limit.

\end{document}